\documentclass[11pt]{article}

\usepackage[T1]{fontenc}
\usepackage[utf8]{inputenc}
\usepackage{lmodern}
\usepackage{authblk}
\usepackage{enumitem}
\usepackage{titletoc}
\usepackage[margin=1in]{geometry}
\usepackage{setspace}
\usepackage{amsmath,amssymb,amsthm}

\usepackage{graphicx}
\usepackage{booktabs}
\usepackage{multirow}
\usepackage{array}
\usepackage{caption}
\usepackage{subcaption}
\usepackage{float}
\usepackage[dvipsnames]{xcolor} 

\definecolor{perRed}{HTML}{EA3221}
\definecolor{perYellow}{HTML}{FFBF00}
\definecolor{perBlue}{HTML}{2274A5}

\usepackage[numbers]{natbib}
\usepackage{hyperref}
\hypersetup{linkcolor  = Gray!45!black, citecolor  = DarkOrchid, urlcolor   = Aquamarine, colorlinks = true}

\newcommand{\acceptedmanuscriptbox}{
    \vskip10mm
    \begin{center}
    \fcolorbox{gray}{white}{
        \begin{minipage}[c]{0.92\textwidth}
        \vskip3mm
        {\Large
        \begin{center}
        \textbf{Accepted Manuscript}\\
        \end{center}
        This is the author’s peer reviewed, accepted manuscript. However, the online version of record will be different from this version once it has been copyedited and typeset.\\[1em]

        The final version is available via DOI:\\
        \url{https://doi.org/10.1016/j.chaos.2026.118251}.\\[1em]

        Please cite the published version:\\[0.5em]

        Millar-Sáez, G., Ormazábal, I., \& Astudillo, H. F. (2026). Parliamentary efficiency under majority and supermajority rules: The role of independent legislators. \textit{Chaos, Solitons \& Fractals}, 208, 118251.\\
        }
        \vskip5mm
        \end{minipage}
    }
    \end{center}
}

\begin{document}

% =========================
% ACCEPTED MANUSCRIPT NOTICE
% =========================
\acceptedmanuscriptbox

\vspace{1em}

% =========================
% MAIN MANUSCRIPT
% =========================

\title{Parliamentary Efficiency under Majority and Supermajority Rules: The Role of Independent Legislators}

\author[1,2]{Gerardo Millar-Sáez}%
\author[1,2,3,4]{Ignacio Ormazábal\thanks{Corresponding author: iormazabal@udec.cl}}
\author[1,2]{Hernán F. Astudillo}
\affil[1]{Departamento de Física, Universidad de Concepción, Concepción, Chile.}
\affil[2]{Grupo de Sistemas Complejos, Universidad de Concepción, Concepción, Chile}
\affil[3]{Computational Research in Social Science Laboratory, School of Engineering and School of Government, Universidad del Desarrollo, Santiago, Chile.}
\affil[4]{Instituto de Data Science, Facultad de Ingeniería, Universidad del Desarrollo, Santiago, Chile.}

\date{}

\maketitle
%==============================================================
\begin{abstract}
Parliaments dominated by two political blocs often face legislative inefficiencies as polarization increases. A central institutional question concerns how majority and supermajority rules interact with parliamentary composition to balance governability and minority protection. This article examines how the inclusion of independent legislators, introduced through sortition, affects collective decision-making under different majority and supermajority quorum requirements.

Using an agent-based model that combines analytical threshold derivations with numerical simulations, we identify four critical thresholds that partition the parameter space into distinct coalition regimes. These regimes range from majority-party dominance to minority veto and, at high levels of diversity, to structural fragmentation in which coordination becomes increasingly difficult.
Parliamentary efficiency emerges from non-linear interactions among quorum thresholds, party-size distribution, and the proportion of independents. Under simple majority, the system exhibits an interior efficiency maximum associated with a transition from unilateral control to pivot-based coalition formation. Under strong supermajorities, however, the same increase in diversity may induce minority veto dynamics or coordination breakdown, significantly reducing legislative performance.

These results show that institutional performance is not an intrinsic property of a particular decision rule but an emergent outcome of the interaction between approval thresholds, parliamentary composition, and the coalition structures they generate.
\end{abstract}

%==============================================================
\section{Introduction}\label{Intro}

In contemporary political landscapes, it is common to see parliaments dominated by two large coalitions or parties \cite{moreira_2006, brewer_2021}, which, in contexts of increasing polarization, tends to translate into severe difficulties for collective decision-making. Evidence shows that divided legislatures are less effective in producing laws \cite{rogers_2005}. Furthermore, depending on the level of polarization and which party is in power, the distribution of legislative seats can significantly reduce legislative efficiency \cite{hicks_2015}.

When preferences are highly fragmented, the result is \textit{legislative gridlock}, a situation in which the government faces persistent obstacles to passing legislative proposals to address urgent social and economic problems \cite{barber_mccarty_2015, bolton_2022}. In this scenario, partisan competition ceases to translate into programmatic alternation and instead generates institutional paralysis.

Legislative gridlock is exacerbated by ideological polarization, understood as the grouping of actors around opposing positions resulting from the alignment of partisan identities \cite{fiorina_2008, mason_2015}. This dynamic is deepened by affective polarization, in which political identity transcends programmatic differences and becomes animosity toward the opponent \cite{iyengar_2019}. Therefore, polarizing rhetoric consolidates negative partisanship, widens social distance, and reduces the willingness to seek cooperative solutions \cite{mccoy_2018}. The result is the paradox in which societies that, despite agreeing on various substantive issues, remain deeply divided by their social identities \cite{mason_2015}. In this context, parliament ceases to operate as a space for deliberation aimed at building agreements. It becomes a stage for persistent confrontation, where intergroup hostility blocks legislative progress and weakens the institutional response to urgent social demands. Polarization thus not only structures political competition, but also directly conditions legislative performance and the social costs associated with government inaction.

In this context, it is crucial to examine how institutional rules that structure parliamentary decision-making can mitigate or deepen polarization. Legislative performance depends not only on the ideological configuration of the actors but also on the rules that determine how their preferences are aggregated. Among these rules, the majority required for the passage of laws occupies a central place. The simple majority, where more than half of the preferences ($50\%+1$) are needed to pass a law, is one of the most widely used rules. However, so-called \textit{supermajorities} are also used, requiring three-fifths, two-thirds, and even five-sixths majorities \cite{laruelle_valenciano_2011}. In particular, the use of supermajorities dates back to 1958 in France and, after successive waves of institutional expansion, has spread widely in different political systems \cite{szentgali_2025}. These rules substantially alter the legislative dynamic by raising the threshold for passing laws.

Regarding which is the most useful or appropriate, there is a long-standing debate in Political science and Social choice theory about the balance between majority rule and minority protection. On the one hand, supermajorities are necessary to safeguard fundamental rights and prevent a drift toward the \textit{tyranny of the majority} \cite{nyirkos_2018, levitsky_ziblatt_2025}. On the other hand, supermajority requirements can lead to greater stagnation and obstruction, as a minority of legislators gains veto power or engages in filibustering, weakening governance, and contributing to the so-called \textit{tyranny of the minority} \cite{levitsky_ziblatt_2024}. In highly polarized contexts, this veto power can intensify paralysis and reduce institutional responsiveness. However, when proposals are successfully approved under these rules, they tend to attract nominally broader coalitions than those formed under simple majority rules \cite{curry_oldham_2025}, which can translate into greater political stability and legitimacy. On the other hand, some argue that the simple majority rule offers greater incentives for negotiation \cite{mcgann_2006}, since no majority is permanent: a minority can form alternative coalitions and reverse undesirable outcomes.

Thus, the problem lies not only in polarization but also in how parliamentarians interact under majority-rule collective decision-making. The rules not only reflect normative conceptions of justice and representation, but also strategically shape the relative power of majorities and minorities in polarized contexts. Consequently, understanding contemporary legislative dynamics requires simultaneously analyzing the structure of preferences and the institutional mechanisms that aggregate them.

In scenarios of parliamentary gridlock, one possible avenue for intervention is the inclusion of independent parliamentarians who can alter the strategic dynamics between dominant parties. Independents can play a relevant role in balancing the dispute between two opposing coalitions, reducing the rigidity of party alignments \cite{ibenskas_etal_2025}. In this context, sortition has historically been proposed as a mechanism for incorporating randomness into decision-making institutions \cite{giuliani_2025}. Then, sortition was a democratic response to inequality and partiality, allowing for the inclusion of randomly selected citizens and limiting institutional capture by organized interests \cite{abizadeh_2021}. Thus, this mechanism allows for the incorporation of legislators who are not aligned with the parties or coalitions that structure political competition. These legislators, not beholden to party structures or traditional electoral incentives, can act as independent agents in evaluating legislative proposals, potentially expanding the space for legislative negotiation.

It has been shown that introducing randomness into social systems can generate benefits in terms of efficiency, stability, adaptation, and cooperation \cite{perc_2006a, perc_2006b, biondo_2013, pluchino_2018, jusup_2022}. In the parliamentary context, numerical simulations suggest that the inclusion of independent legislators selected by lottery can improve parliamentary performance \cite{pluchino_2011a, caserta_2021, sotgiu_2025}. Likewise, formal models examine how sortition can enhance legislative efficiency across different institutional configurations \cite{bruzzone_2025}.

Despite these advances, the literature has tended to analyze the effects of supermajority rules and the effects of including independent legislators separately. Much less explored has been the interaction between the two mechanisms: how does legislative performance vary when the incorporation of independents is combined with different approval thresholds? In polarized contexts dominated by two parties or coalitions, this interaction is crucial to understanding whether majority rule reinforces gridlock or instead reconfigures coalition structures in ways that either facilitate negotiation, empower minority veto power, or induce fragmentation.

To address this question, the article employs the agent-based model proposed by Pluchino et al. \cite{pluchino_2011a} to analyze the collective behavior of a parliament composed of a majority party, a minority party, and independent legislators under alternative majority rules. The results identify five distinct phases, determined by the coalition structures that emerge among these actors. These phases range from unilateral dominance by the majority party, to configurations requiring cross-party support, to regimes in which independents act as strategic pivots, and finally to contexts in which a high proportion of independents reduces the system’s capacity for legislative coordination.

When evaluating the different majority rules, the results show that their effect is neither linear nor uniform. At a fixed percentage of majority party members, an increase in the quorum reduces the proportion of independents needed to achieve maximum efficiency levels. However, more demanding majorities also decrease the number of proposals passed and, in certain scenarios, the aggregate social welfare. In highly polarized contexts, the simple majority rule is the only one that guarantees sustained efficiency, suggesting that the presence of independents under moderate thresholds favors more inclusive negotiation dynamics. In contrast, when the majority party holds a dominant position, higher quorums can improve efficiency and limit the risks associated with the tyranny of the majority, albeit at a cost to legislative output or programmatic diversity.

Taken together, this study integrates two strands of the literature that have largely evolved separately: the institutional analysis of quorum rules and the study of independent representation mechanisms. By examining their interaction within a unified modeling framework, the article shows that legislative performance cannot be attributed solely to majority rule or to the mere inclusion of independent legislators. Instead, it emerges from the configuration of coalition structures shaped jointly by approval thresholds and parliamentary composition. By linking analytically derived thresholds to simulated regime transitions, the model provides a systematic framework for understanding how different institutional arrangements produce distinct patterns of dominance, veto, negotiation, or fragmentation in polarized political systems.

The article is organized as follows: Section \ref{model} presents the model, the system metrics, and the characteristic thresholds that determine collective dynamics. Then, in Section \ref{results}, we present the results of the numerical simulation for the different majority rules. Finally, in the Section \ref{remarks}, we discuss our results and conclusions.

%==============================================================
\section{Parliamentary dynamics model}\label{model}

The agent-based model proposed by Pluchino et al. \cite{pluchino_2011a} simulates the dynamics of legislative proposal and voting in a unicameral parliament composed of $N$ legislators. A fraction of them are elected by vote and affiliated with political parties, while the remainder are selected by sortition. We refer to these latter legislators as independent, since they are not subject to party discipline and act autonomously in evaluating legislative proposals.

Although the model allows for multiple parties, in this work we consider only two: a majority party ($P_M$) and a minority party ($P_m$). Let $p$ denote the proportion of party-affiliated legislators who belong to the majority party. Given $N_{ind}$ independent legislators, the remaining $N - N_{ind}$ legislators are affiliated with political parties and are distributed as $N_M = (N - N_{ind}) p$ and $N_m = (N - N_{ind}) (1 - p)$.

Each legislator $i$ occupies a position $l_i(x_i, y_i)$ on a Cartesian plane normalized on $[-1,1]$, representing the interests that guide their behavior. The $x$ axis denotes the personal gain associated with a legislative proposal, while the $y$ axis represents its social gain. Independent legislators are randomly distributed on the plane with a uniform distribution. In contrast, legislators affiliated with a party $k$ are located within a circle $C_k(x_k, y_k)$ of radius $r_k$. The center of the circle represents the average behavior of its members, based on the party's values and principles, and the circumference indicates the party's tolerance for internal dissent. The center of the circle is assigned using a uniform distribution on the plane.

The model's dynamics evolve over a legislative $L$ composed of $N_a$ legislative acts. During each act, a legislator $i$ is randomly selected to make a legislative proposal $a_n(x_n^*, y_n^i)$. The coordinate $y_n^i$ coincides with the proponent's position $y_i$ and represents the proposal's social gain, assumed to be commonly perceived by all legislators. In contrast, the coordinate $x_n^*$ is uniformly distributed over $[-1,1]$ and represents the individual gain each legislator obtains when evaluating the proposal.

%------------------------------
 \begin{figure}
     \centering
     \includegraphics[width=0.45\textwidth]{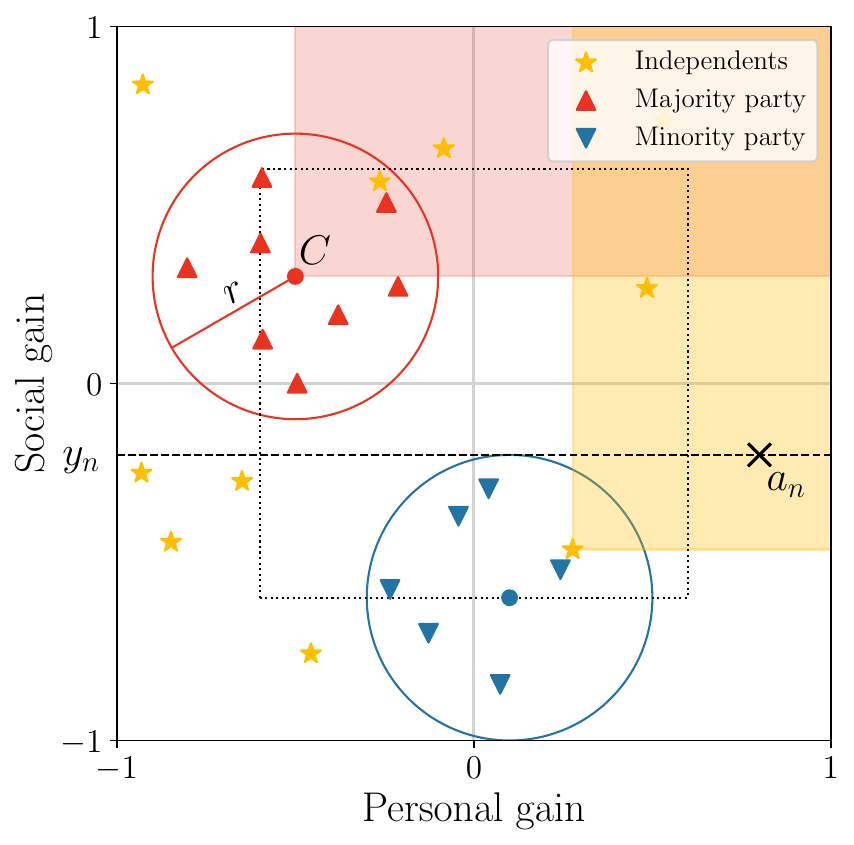}
     \caption[Figure 1]{\label{fig_1} Distribution of agents and acceptance windows in the Cartesian plane. Example distribution of independent legislators (\textcolor{perYellow}{$\bigstar$}) and political party affiliates (\textcolor{perRed}{$\blacktriangle$} and \textcolor{perBlue}{$\blacktriangledown$}) within the circle defined by the center $C$ ($\bullet$) and the tolerance radius $r$. The square area enclosed by the dotted line corresponds to the possible locations of the party centers. The shaded areas represent the acceptance windows for the majority party and an independent legislator. The legislative proposal $a_n$ ($\times$) lies in the line $y_n$, and their x position depends on the agent that votes for them. In this figure, all legislators and parties below the line $y_n$ approve the proposal $a_n$ because their acceptance window contains the proposal's position.}
 \end{figure}
%------------------------------

Next, each legislator $j$ decides their vote using an acceptance window. For an independent legislator located at $(x_j, y_j)$, a proposal is accepted if $x_n^* \ge x_j$ and $y_n^i \ge y_j$, defining a rectangle composed of all points greater than or equal to their position in the plane. In the case of legislators affiliated with a party, the acceptance window is determined by the party center $(x_k, y_k)$, which implies perfect party discipline in evaluating external proposals. Figure \ref{fig_1}  schematically shows the positions of the parties, each independent legislator, and the acceptance window for each party.

%------------------------------
 \begin{figure}
     \centering
     \includegraphics[width=0.50\textwidth]{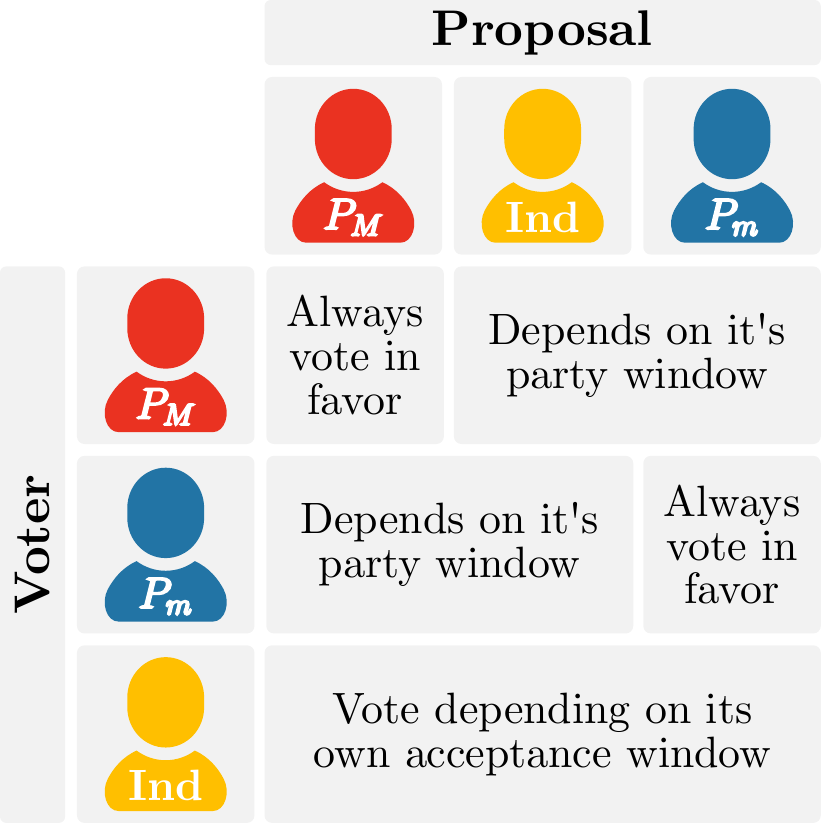}
     \caption[Figure 2]{\label{fig_2} Voting rules are based on the affiliation of the legislator submitting the proposal. Each independent legislator considers voting in favor of the proposal if it is within his or her window of acceptance, in addition to voting in favor of the proposals he or she makes. On the other hand, a party's legislator will vote in favor of a proposal if the proposal is within the party's acceptance window or if the proponent belongs to the same party.}
 \end{figure}
%------------------------------

The voting is structured as follows: the proposing legislator always votes in favor of their own initiative; independent legislators vote exclusively according to their acceptance window; party legislators automatically vote in favor if the proposal comes from a member of their party and, otherwise, decide according to the acceptance window defined by the party center. Figure \ref{fig_2} shows a schematic summary of the voting rules. A proposal is considered approved if the number of favorable votes equals or exceeds the threshold $qN + 1$, where $q$ represents the quorum required according to the majority rule necessary to pass a proposal (e.g., $1/2$, $4/7$, $3/5$, or $2/3$).

At the end of the $N_a$ legislative acts, a legislature $L$ is completed, and parliamentary performance is evaluated. Three metrics are used. The first is the percentage of accepted proposals, defined as 

    \begin{equation} \label{eq_1}
        N_{\%acc}(L) = \frac{N_{acc}}{N_a} \cdot100\%,
    \end{equation} 

\noindent where $N_{acc}$ corresponds to the number of approved proposals in the legislature $L$. The second metric is the average social welfare of accepted proposals, defined as 

    \begin{equation} \label{eq_2}
        Y(L) = \frac{1}{N_{acc}} \sum_{n=1}^{N_{acc}} y_n,
    \end{equation}
    
\noindent with a range of $[-1,1]$. Since passing many laws with low social welfare or few laws with high social welfare does not necessarily reflect good performance, a third parliamentary efficiency metric is introduced \cite{pluchino_2011a}, defined as 

    \begin{equation} \label{eq_3}
        E\!f\!f(L) = N_{\%acc}(L)\cdot Y(L),
    \end{equation}
    
\noindent which integrates the quantity and quality of legislative output and can range from $-100$ to $100$.

\begin{figure}
     \centering
     \includegraphics[width=1.0\textwidth]{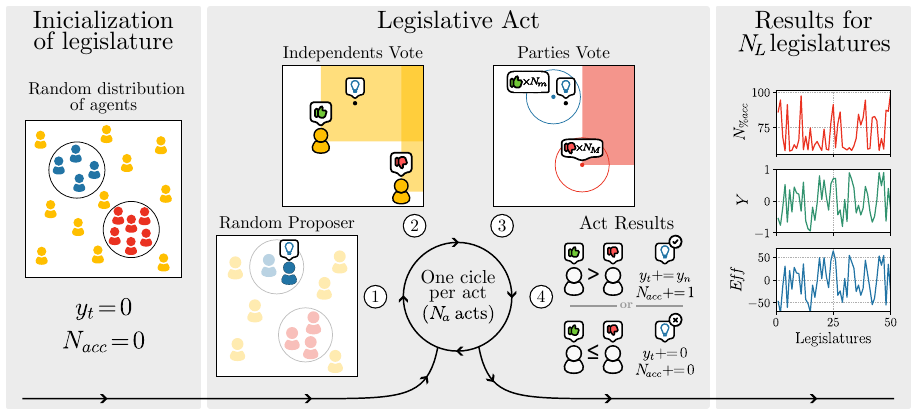}
     \caption[Figure 3]{\label{fig_3} Schematic visualization of the dynamics of the model. A legislature starts with a random distribution of independent agents and agents belonging to parties. Then, a legislative act begins, selecting a random agent and proposing a law that will be voted on by all the agents in the system, following the voting rules. After verifying whether the law is approved or rejected, another legislative act is carried out until there is a complete legislature. Finally, the metrics used to evaluate the parliament's performance are calculated.}
 \end{figure}

Figure \ref{fig_3} presents a schematic of the model's dynamics following the V-ODD protocol \cite{szangolies_2024}, while a complete description according to the ODD protocol \cite{grimm_2020} is included in the supplementary material.

%------------------------------
%------------------------------

A key aspect for understanding the model's dynamics is the number of independent legislators in the system. Variations in $N_{ind}$ alter the relative majorities required to pass a legislative proposal and, therefore, modify the microscopic structure of support among groups. These changes are reflected in the system's macroscopic states, such as the percentage of proposals passed, average social welfare, and parliamentary efficiency, allowing the identification of regime transitions and the construction of a phase diagram characterizing collective behavior.

First, when the number of independents is sufficiently low, the majority party can achieve the required quorum to pass laws on its own if $N_M \ge qN + 1$. Since $N_M = (N - N_{ind})p$, it is possible to determine the critical number of independents beyond which the majority party loses the ability to pass laws without external support. This first threshold is defined as:

\begin{align}
    \label{eq_a}
    \text{thr. } A: \quad N_{ind}^A = \frac{N(p-q) - 1}{p}.
\end{align}

For $N_{ind} > N_{ind}^A$, the majority party necessarily requires the support of other groups to reach a quorum. Since independent legislators are evenly distributed on the Cartesian plane, on average, one-quarter of them vote in favor of a typical proposal \cite{pluchino_2011a}. Consequently, the majority party can regain the ability to pass laws if: $N_M + \frac{N_{ind}}{4} \ge qN + 1$. From this condition, a second threshold is obtained:

\begin{align}
    \label{eq_b}
    \text{thr. }B: \quad N_{ind}^B = \frac{N(p - q) - 1}{p - 0.25},
\end{align}

This expression is analogous to the so-called "golden rule of efficiency" reported by Pluchino et al. \cite{pluchino_2011a}, since numerical simulations show that around this threshold, maximum efficiency is observed when incorporating independent variables.

A third possible configuration for reaching a quorum occurs when both parties vote jointly in favor of a proposal, that is, when: $N_M + N_m \ge qN + 1$. Using the identity $N = N_M + N_m + N_{ind}$, we obtain the third threshold:

\begin{align}
    \label{eq_c}
    \text{thr. }C: \quad N_{ind}^C = N(1 - q) - 1.
\end{align}

Finally, there is a fourth regime in which approval requires the joint support of both parties and a fourth independent vote that, on average, votes in favor. In this case, the condition $N_M + N_m + \frac{N_{ind}}{4} \ge qN + 1$ leads to:

\begin{align}
    \label{eq_d}
    \text{thr. }D: \quad N_{ind}^D = \frac{4}{3}N(1 - q) - \frac{4}{3}.
\end{align}

The thresholds $A$, $B$, $C$, and $D$ delimit regions with distinct microscopic dynamics that give rise to collective phases visible at the macroscopic level. Essentially, each threshold marks a transition in the structure of support required to pass a proposal: from the exclusive dominance of the majority party, through configurations in which independents act as strategic pivots, to scenarios in which coordination among multiple groups becomes indispensable. These analytical results allow us to anticipate how different quorums modify the architecture of legislative coalitions and provide the conceptual basis for interpreting the phase diagram presented in the following subsection. For a detailed derivation of the analytical expressions, see the supplementary material.

%==============================================================
\section{Results and discussion}\label{results}

%===

We performed all numerical simulations using the following fixed parameters: a parliament size of $N=1000$ legislators, party radius $r_k=r=0.2$, and $N_a=1000$ legislative acts per legislature. Since the model scales with the number of agents, we selected $N=1000$ as a compromise between computational cost and statistical stability. A discussion of scaling properties is provided in the supplementary material.

The analyses systematically explore three key parameters of the model: the proportion of legislators belonging to the majority party, $p$; the number of independent legislators, $N_{ind}$; and the required quorum $q$. We vary the majority party proportion over $p \in [0.5,1]$ in discrete increments of $0.5$, covering configurations from polarized settings ($p$ close to $0.5$) to dominant-party regimes ($p$ approaching $1$). 

The number of independent legislators varies from $0$ to $1000$ in discrete increments of $2$. We consider four institutional quorums corresponding to different majority rules: $q=1/2$, $4/7$, $3/5$, and $2/3$. For each configuration $(p, N_{ind}, q)$, we perform $1000$ independent realizations, corresponding to $1000$ legislatures $L$, and compute the percentage of accepted proposals, the mean social welfare, and the overall parliamentary efficiency.

To characterize the collective behavior of the system, we first compute the macroscopic performance metrics for each parameter combination. We then numerically identify the transition points separating different regimes of coalition formation and compare them with the analytical thresholds derived in Section \ref{model}. Using these coordinates in the $(p, N_{ind})$ plane, we construct the phase diagrams that reveal the five qualitatively distinct regimes of parliamentary dynamics. These phases correspond to different coalition structures required to reach the institutional quorum under the selected parameter values. All reported curves and phase boundaries correspond to ensemble averages over the $1000$ independent legislatures, ensuring statistical robustness of the results.

To understand how the interaction between party structure and decision-making rules translates into different parliamentary operating regimes, we organized the results into three subsections. First, we analyzed the microscopic dynamics of the model in a polarized scenario under simple majority rule. Second, we examined the same polarized context under a strong supermajority to identify how the phase architecture changes as the institutional threshold becomes more demanding. Finally, we presented an aggregate analysis that systematizes the combined effect of the majority party size and quorum on parliamentary efficiency.

%===
\subsection{Polarized scenario under simple majority:\texorpdfstring{$p=0.55$, $q=1/2$}{p=0.55, q=1/2}}

\begin{figure}
     \centering
     \includegraphics[width=1.0\textwidth]{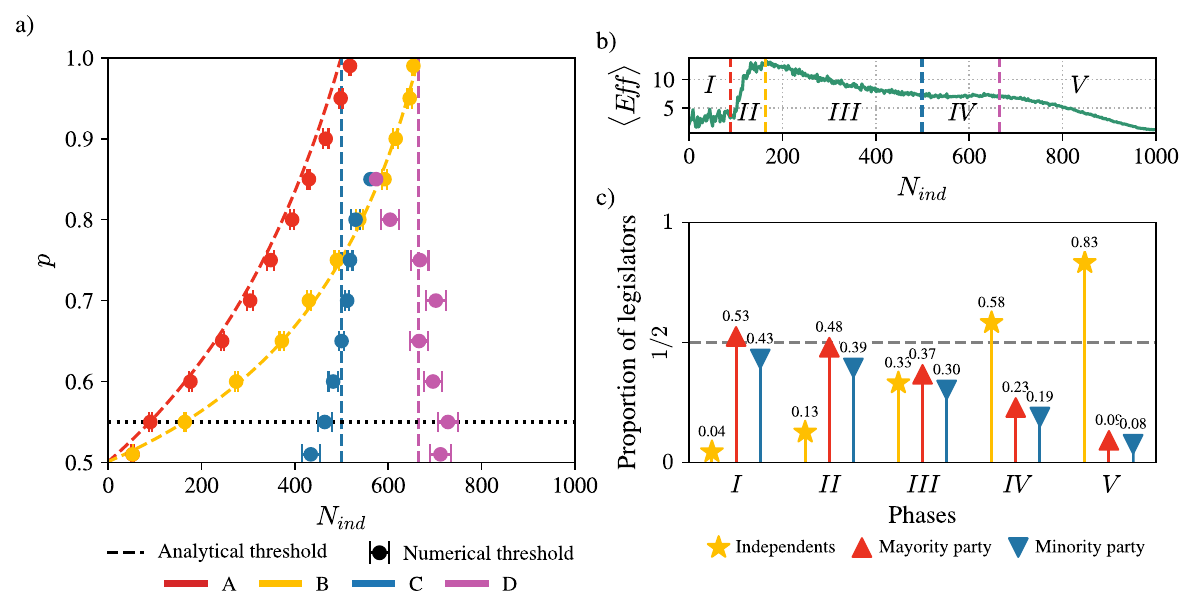}
     \caption[Figure 4]{\label{fig_4} Parliamentary dynamics under simple majority in a polarized scenario ($p=0.55$, $q=1/2$). a) Phase diagram showing the five phases ($I$–$V$) determined by the analytical thresholds $A$, $B$, $C$, and $D$. The horizontal dashed line indicates the value of $p=0.55$ used in this scenario. Phases correspond to distinct coalition structures required to reach the quorum. b) Average parliamentary efficiency $\langle E\!f\!f \rangle$ as a function of the number of independent legislators $N_{ind}$. Vertical dashed lines mark the threshold values separating phases. c) Representative composition of the parliament in each phase, showing the proportion of majority party, minority party, and independent legislators. The dashed horizontal line indicates the quorum required for legislative proposal approval ($50\%+1$). Simulation parameters: $N=1000$, $N_a=1000$, $r=0.2$, and $1000$ legislative terms averaged for each data point. The figure shows that increasing the number of independent legislators leads to successive regime shifts in the coalition structure needed to reach the approval quorum. Parliamentary efficiency peaks between Phases $II$ and $III$, where independents improve the social welfare of approved proposals without yet causing coordination breakdown.}
 \end{figure}

Figure \ref{fig_4} shows the model's dynamics in a polarized scenario, where the majority party holds $55\%$ of the party-affiliated legislators, and the parliament approves proposals by simple majority. Figure \ref{fig_4} a) presents the corresponding phase diagram, where the horizontal line associated with $p=0.55$ successively intersects the analytical thresholds $A$, $B$, $C$, and $D$, generating five distinct phases ($I$–$V$) as the number of independent legislators ($N_{ind}$) increases. These phases represent qualitative changes in the minimal coalition required to approve proposals, reflecting shifts in the system’s underlying coalition structure.

Figure \ref{fig_4} b) shows the average parliamentary efficiency ($E\!f\!f$) as a function of $N_{ind}$, indicating that the system exhibits a maximum efficiency. In Phase $I$, with very few independents, the majority party surpasses the simple majority threshold on its own and dominates legislative approval. While the approval rate is high, efficiency does not reach its maximum, suggesting that legislative output primarily reflects the dominant bloc's agenda. Upon entering Phases $II$ and $III$, the majority loses the ability to pass laws without additional support, and independents assume a pivotal role. This intermediate stage represents peak efficiency: the diversity introduced by independents raises the average social welfare without destroying legislative coordination. From Phase $IV$ onward, when independents become the majority, ideological fragmentation erodes the capacity to form stable coalitions, and efficiency progressively declines until Phase $V$, when the overrepresentation of independents leads to poor coordination and lower overall performance.

Figure \ref{fig_4} c) allows for the interpretation of these phases in terms of parliamentary composition. In Phase $I$, the majority party holds approximately $53\%$ of the party-affiliated legislators, exceeding the approval threshold and operating under a logic akin to majority rule. In Phases $II$ and $III$, its share falls below $50\%$, forcing it to negotiate with independents or the minority party. In Phases $IV$ and $V$, independents significantly outnumber traditional parties, creating a system where diversity prevails over coordination. Overall, this scenario demonstrates that under polarization, a simple majority can reach an optimal point in which independents serve as a corrective mechanism against rigid bipartisanship, fostering a balance between the quantity and quality of legislation.

%------------------------------
%------------------------------

%===
\subsection{Polarized scenario under a strong supermajority:\texorpdfstring{$p=0.55$, $q=2/3$}{p=0.55, q=2/3}}

\begin{figure}
     \centering
     \includegraphics[width=1.0\textwidth]{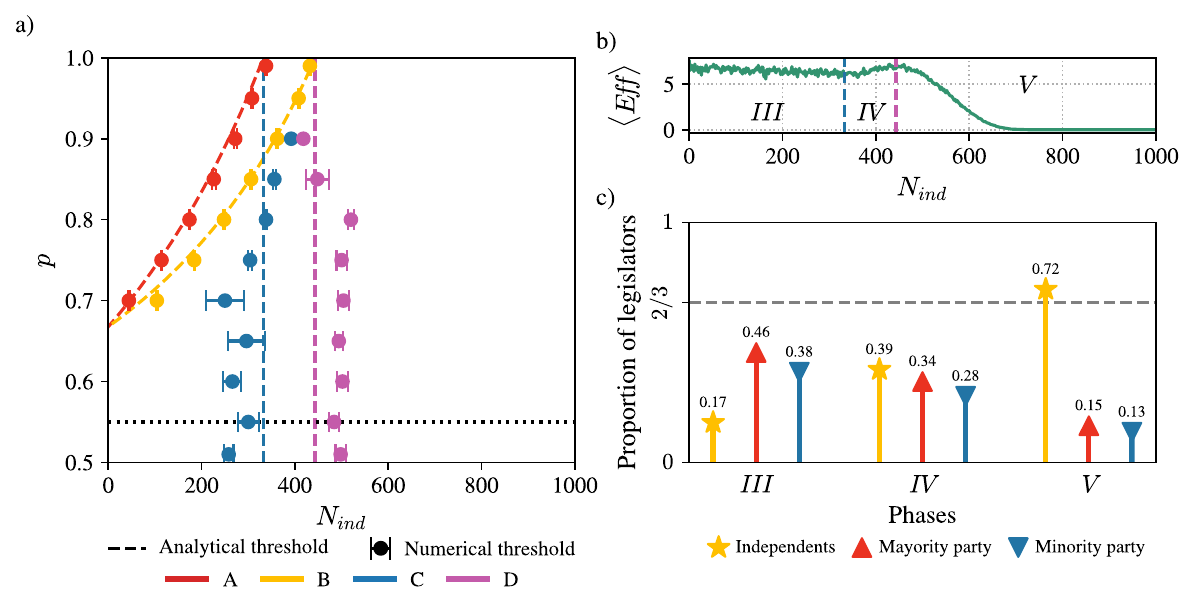}
     \caption[Figure 5]{\label{fig_5} Parliamentary dynamics under strong supermajority in a polarized scenario ($p=0.55$, $q=2/3$). a) Phase diagram illustrating the phases intersected by the horizontal line corresponding to $p=0.55$. In contrast to the simple majority case, the system operates only in Phases $III$–$V$ due to the higher approval threshold. Phases correspond to distinct coalition structures required to reach the quorum. b) Average parliamentary efficiency $\langle Eff \rangle$ as a function of the number of independent legislators $N_{ind}$. Vertical dashed lines denote the analytical thresholds separating phases. c) Representative parliamentary composition in each phase, displaying the proportions of majority party, minority party, and independent legislators. The dashed horizontal line indicates the required two-thirds quorum for legislative proposal approval. Simulation parameters: $N=1000$, $N_a=1000$, $r=0.2$, and $1000$ legislative terms averaged per configuration. The figure shows that under a two-thirds quorum, the system structurally needs broad multi-party coalitions, since no single party can reach the approval quorum on its own. Parliamentary efficiency remains relatively stable while a two-party structure persists (Phases $III$–$IV$), but collapses once independents dominate the chamber (Phase $V$), where forming a supermajority becomes increasingly unlikely.}

 \end{figure}

Figure \ref{fig_5} presents the same polarized scenario ($p=0.55$) under a two-thirds quorum. The phase diagram in Figure \ref{fig_5} a) reveals a substantial structural difference compared to the simple majority case: the line corresponding to $p=0.55$ does not pass through all five phases, but begins directly in Phase $III$. The reason is that no party individually reaches the required threshold. Unlike the simple majority case, the system operates structurally in configurations in which approval depends on broad coalitions exceeding $66.7\%$ of the quorum.

Figure \ref{fig_5} b) shows that average efficiency remains relatively stable in Phases $III$ and $IV$, but collapses abruptly upon entering Phase $V$. Unlike the simple majority case, there is no pronounced efficiency peak associated with the transition from party dominance to independent pivot. Instead, the system permanently relies on broad multi-party agreements. As long as the parliamentary composition retains a significant two-party structure (Phases $III$ and $IV$), it is still possible to coordinate sufficient support. However, when the number of independents exceeds a certain threshold, the probability of forming a supermajority becomes extremely low, leading to a regime of structural paralysis in which the approval rate falls sharply, and efficiency converges to near zero.

Figure \ref{fig_5} c) allows for a clearer understanding of the microscopic dynamics underlying this behavior. In Phase $III$, the majority party maintains a relatively dominant position ($ \approx46\% $), but the two-thirds threshold prevents it from exercising its majority without the minority party's simultaneous support. At this point, a scenario akin to the tyranny of the minority emerges: the minority bloc acquires structural veto power, as its refusal to cooperate prevents the required quorum from being reached. In Phase $IV$, the growth of independent candidates further complicates the formation of broad coalitions, increasing the reliance on cross-party agreements. Finally, in Phase $V$, when independent candidates far outnumber traditional parties, the dynamics shift from strategic veto power to structural fragmentation. In this stage, the dispersion of preferences makes it virtually impossible to coordinate stable supermajorities, consolidating a regime of institutional paralysis distinct from the minority veto power observed in previous phases.

The comparison between the polarized scenario under simple majority ($q = 1/2$) and the same configuration under a strong supermajority ($q = 2/3$) shows that the effect of independent legislators depends critically on the prevailing quorum and the type of microscopic regime it induces. With a simple majority, independents can act as pivots, energizing negotiations and maximizing efficiency in polarized contexts, shifting the system from majority dominance toward inter-party cooperation. Under a strong supermajority, however, the same increase in diversity interacts with the institutional threshold differently. In Phase $III$, the system can become trapped in a minority veto regime, in which the majority party cannot exercise its majority without the minority's consent. In later stages, as the number of independents grows, the paralysis no longer stems from a strategic veto but from structural fragmentation, that is, the inability to coordinate sufficiently broad coalitions. Consequently, the relationship between parliamentary composition and institutional performance is neither linear nor universal: it depends on the interaction between structural polarization, collective decision-making rules, and the coalition architecture that these rules induce.

%------------------------------
%------------------------------

The two scenarios above clearly demonstrate how the quorum alters the microscopic dynamics of coalitions in a polarized context. However, these results correspond to a fixed majority party size. To more generally assess how parliamentary efficiency depends on both quorum and party structure, a more systematic perspective is needed. In the following subsection, we analyze the system's aggregate metrics, examining how the point of maximum efficiency and associated performance vary as we modify both the institutional threshold and the proportion of the majority party.

%===
\subsection{Parliamentary efficiency and the effects of quorum}

\begin{figure}
    \centering
    \includegraphics[width=1.0\linewidth]{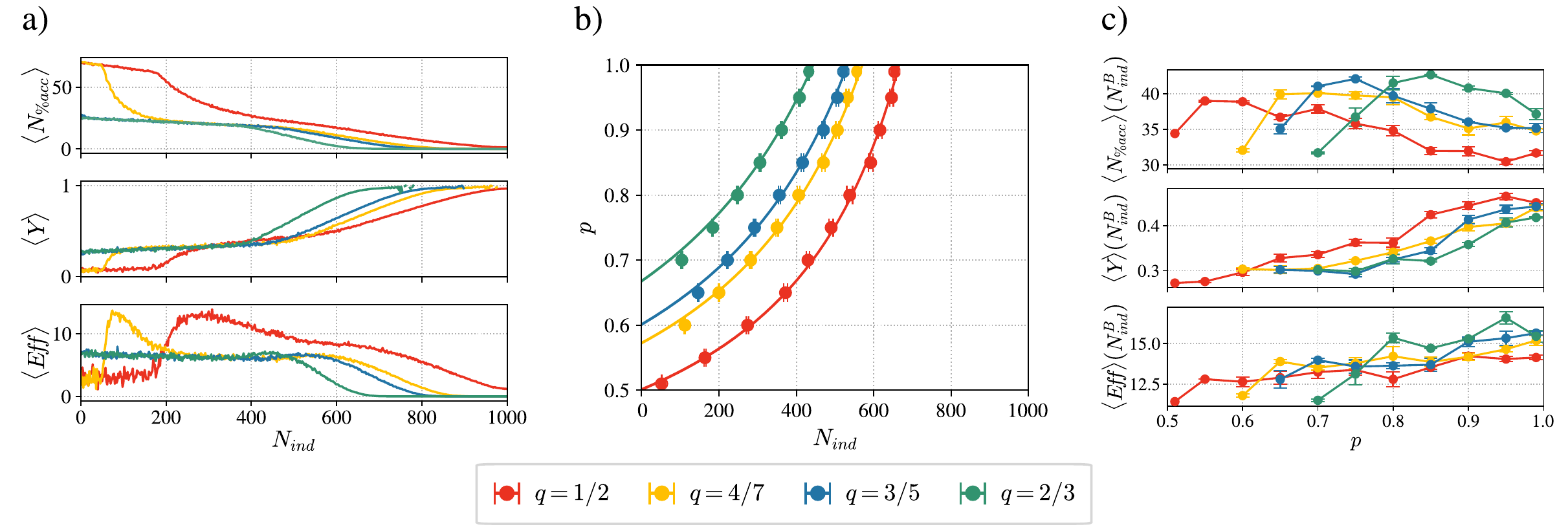}
    \caption[Figure 6]{ \label{fig_6} Aggregate effects of quorum rules on parliamentary performance. a) Variation of macroscopic metrics—percentage of accepted proposals, average social gain, and overall efficiency—as a function of the number of independent legislators $N_{ind}$ for different quorum values $q$ and $p=0.6$. b) Position of the efficiency-maximizing point (threshold B) in the $(p, N_{ind})$ plane for different quorum requirements, showing how the optimal proportion of independents shifts with the size of the majority party. c) Values of the macroscopic metrics evaluated at the efficiency-maximizing configuration ($N_{ind}^{B}$) for each quorum, illustrating how institutional thresholds shape performance at the optimum. Simulation parameters: $N=1000$, $N_a=1000$, $r=0.2$, and $1000$ legislative terms averaged for each parameter configuration. Overall, the figure shows that parliamentary efficiency emerges from a balance between party coordination and ideological diversity. The location and performance of the efficiency maximum depend jointly on the approval quorum and the distribution of power among parties, indicating that institutional rules and parliamentary composition interact to determine the system's optimal configuration.}

\end{figure}

Figure \ref{fig_6} presents the system's behavior under different quorum rules and allows us to identify how the inclusion of independent legislators modifies parliamentary dynamics. We organized the analysis into three levels: (i) the existence of a maximum efficiency, (ii) the shift of that maximum when the party structure changes, and (iii) the system's performance at that optimal point.

Figure \ref{fig_6} a) shows the variation of macroscopic metrics as a function of the number of independent legislators ($N_{ind}$) for different quorum values ($q$). We observe a robust pattern: as $N_{ind}$ increases, the percentage of accepted proposals ($N_{\%acc}$) decreases monotonically, while the average social welfare ($\langle Y \rangle$) increases. This behavior reflects a structural trade-off in the model. The introduction of independents increases the diversity of ideological positions and, therefore, raises the average standard of proposal acceptance (higher social welfare). However, this same diversity hinders legislative coordination and reduces the approval rate. As a result, average parliamentary efficiency exhibits an inner maximum for intermediate values of $N_{ind}$. At this maximum, there is an optimal number of independents in parliament, where laws pass and, at the same time, have a higher average social welfare, achieving an equilibrium. The quorum effect clearly differentiates, as it generates a systematic shift in this equilibrium. As the quorum increases, the efficiency point shifts to the left, reducing the range of maximum efficiency until it disappears at high quorums. On the other hand, under more stringent rules (e.g., $q=2/3$), both the decline in approval rates and the decline in efficiency fall more rapidly as the number of legislators in the system decreases. In contrast, for less stringent rules, such as simple majority ($q=1/2$), a wider and more robust maximum is maintained. Furthermore, the decline in efficiency is slower as the number of independent legislators increases. These results suggest that, in contexts with greater diversity of preferences, high thresholds transform heterogeneity into paralysis. At the same time, less stringent rules allow the diversity introduced by independent legislators to translate into effective negotiation.

For each combination of $p$ and $q$, we define the optimal number of independents as 

\begin{equation*}
N_{ind}^{opt}(p,q) = \arg\max_{N_{ind}} \langle Eff(N_{ind};p,q) \rangle,
\end{equation*}

\noindent that is, the value of $N_{ind}$ that maximizes the system's average efficiency.

Figure \ref{fig_6} b) shows how $N_{ind}^{B}$ varies as a function of the majority party's percentage $p$, for each quorum rule. We observe two structural regularities. First, as $p$ increases, the optimal number of independents decreases. When the majority party holds a larger proportion of party-affiliated legislators, legislative coordination becomes structurally simpler. The dominant party is closer to the approval threshold and requires fewer additional votes to reach a quorum. Consequently, the pivotal role of independents is reduced. Second, for the same value of $p$, the most demanding quorum rules reach their maximum efficiency with fewer independents. High thresholds restrict the set of eligible proposals and penalize the dispersion of preferences; Therefore, the system maximizes efficiency in configurations with less parliamentary heterogeneity.

However, this result has a relevant normative implication. When $p$ is high, the shift of the optimum towards low $N_{ind}$ values not only reflects greater ease of coordination but also a greater capacity of the majority party to impose its agenda. In these scenarios, independents act as marginal pivots that allow the required quorum to be reached, but they do not necessarily introduce a substantive counterweight. On the contrary, they can reinforce the decision-making capacity of the dominant bloc. Thus, the point of maximum efficiency can coexist with dynamics close to the tyranny of the majority.

Likewise, for low values of $p$, where scenarios are close to a competitive, highly polarized two-party system, certain high quorums do not achieve maximum efficiency. This result indicates that under polarized configurations, strict rules can prevent the very existence of an efficient equilibrium, reinforcing the idea that the interaction between polarization and supermajorities can induce legislative gridlock.

Figure \ref{fig_6} c) shows the metrics evaluated at the point of maximum efficiency $N_{ind}^{opt}(p,q)$. This analysis allows examination of institutional performance when the system operates at its most efficient configuration for each parameter combination. We observe that the average social welfare, $\langle Y \rangle$, increases with the size of the majority party across all quorums. Part of this effect is explained by the fact that a larger $p$ implies that the majority party proposes a greater fraction of legislative initiatives, which more frequently aligns the legislative agenda with its median position in the ideological space. Consequently, the observed efficiency reflects not only an improvement in deliberative quality but also greater coherence between the power structure and legislative output.

The percentage of accepted proposals $\langle N_{\%acc} \rangle$, however, remains conditioned by the institutional quorum. In simple-majority scenarios, the system maintains sustained approval levels even in polarized contexts. Conversely, with high quorum requirements, the approval rate can collapse when $p$ is low, demonstrating that the institutional requirement amplifies existing fragmentation.

Thus, on the one hand, in highly polarized scenarios (low $p$), a simple majority is the only rule that guarantees sustained efficiency levels. Supermajorities, on the other hand, can induce dynamics bordering on minority tyranny, in which the requirement for broad support systematically blocks the passage of initiatives. On the other hand, in scenarios with strong dominance of the majority party (high $p$), more demanding quorum requirements can achieve comparable or even superior efficiencies. However, such efficiency may be associated with configurations in which the legislative agenda predominantly reflects the preferences of the dominant bloc, and in which independents operate as marginal facilitators rather than corrective actors.

Taken together, the figure shows that parliamentary efficiency does not depend solely on the proportion of independents, but rather on the interaction between parliament's composition and the quorums required to pass laws. Figure \ref{fig_6} a) reveals a structural trade-off between legislative output and social welfare that generates efficiency maxima; Figure \ref{fig_6} b) shows that the location of these maxima systematically depends on the size of the majority party and the approval threshold; and Figure \ref{fig_6} c) indicates that performance at the optimum can take on very different institutional configurations.

In particular, the system can achieve high efficiency both in contexts where a simple majority allows the diversity introduced by independents to translate into effective negotiation, and in scenarios where the dominance of the majority party, combined with a small number of pivotal independents, consolidates the decision-making capacity of the dominant bloc. Conversely, high quorums can serve as protective mechanisms against large majorities, but in polarized contexts, they can induce dynamics akin to minority tyranny by preventing the formation of viable coalitions. Thus, efficiency is not a univocal indicator of democratic equilibrium, but the contingent result of the interaction between decision rules and power distribution.

%------------------------------
%------------------------------

Taken together, the results show that legislative performance cannot be understood solely in terms of majority size or quorum rules, but depends on their interaction and the role of independent legislators across regimes. Microscopic analysis using the phase diagram allows identification of qualitative transitions in coalition structure, while aggregate metrics reveal how these transitions translate into observable changes in parliamentary efficiency. Thus, microscopic coalition configurations give rise to distinct macroscopic patterns of parliamentary performance.

Under a simple majority rule, the system can encompass a wide range of configurations, from the dominance of the majority party to regimes in which independents serve as pivotal centers. In this context, a maximum efficiency emerges associated with an intermediate balance between party coordination and ideological diversity. By contrast, under strong supermajorities, the system can become trapped in regimes where the minority party wields structural veto power, creating scenarios bordering on minority tyranny, or evolve toward situations of structural fragmentation where the dispersion of preferences makes it impossible to reach broad agreements. These contrasting patterns highlight that institutional thresholds reshape not only approval probabilities but the very structure of feasible coalitions.

Overall, these findings indicate that parliamentary efficiency is not an inherent property of a specific majority rule, but an emergent outcome shaped by the interaction between institutional thresholds and parliamentary composition. The inclusion of independent legislators may foster negotiated coalition equilibria under moderate approval rules, reinforce dominant agendas when majority concentration is high, or exacerbate coordination failures under demanding supermajority requirements. The model therefore shows that the classic tension between majority dominance and minority obstruction cannot be resolved by selecting a particular quorum in isolation. Instead, it depends critically on the interaction between institutional thresholds, parliamentary composition, and the presence of independent legislators, which together shape the structure of feasible coalitions.

%==============================================================
\section{Concluding Remarks}\label{remarks}

The results provide relevant implications for the institutional design of parliaments in contemporary contexts marked by polarization and fragmentation. Legislative performance, as shown by the model, does not depend exclusively on the type of majority rule adopted, but on how institutional thresholds interact with party size distribution and the presence of independent legislators. The relationship between quorum and efficiency is therefore neither linear nor universal; it is contingent upon the political configuration in which it operates.

In polarized scenarios, simple majority rules allow independent legislators to serve as pivotal actors, softening rigid bipartisanship, and yielding configurations in which diversity translates into negotiated, efficient coalition structures. Under strong supermajority requirements, however, the same context may give rise either to minority veto regimes, in which the majority party cannot exercise its majority without the consent of the minority party, or to structural fragmentation, in which assembling sufficiently large coalitions becomes increasingly unlikely.

The analysis identifies three structural mechanisms shaping legislative performance: (i) majority dominance under flexible thresholds, (ii) minority veto induced by demanding approval rules, and (iii) coordination failure when parliamentary diversity exceeds institutional capacity. Importantly, similar levels of efficiency may emerge from qualitatively different coalition regimes, indicating that aggregate performance metrics alone do not fully capture the underlying political dynamics.

From a normative perspective, these findings suggest that reforms concerning quorum rules or independent representation cannot be evaluated in isolation. No rule is universally superior: supermajorities may constrain dominant majorities but risk empowering organized minorities, while simple majorities may sustain governability in competitive contexts yet consolidate dominance under high party concentration.

Ultimately, the classic dilemma between majority and minority tyranny cannot be resolved by adopting a quorum requirement in the abstract. Rather, it depends on the dynamic interaction between institutional rules, parliamentary composition, and the coalition structures they generate. By conceptualizing legislative efficiency as an emergent property of this interaction, the article offers a systematic framework for assessing parliamentary reforms in polarized political systems.

%==============================================================
\section{CRediT authorship contribution statement}

{\bf Gerardo Millar-Sáez:} Conceptualization, Methodology, Software, Formal analysis, Investigation, Writing - original draft, Writing - review \& editing. {\bf Ignacio Ormazábal:} Conceptualization, Methodology, Formal analysis, Investigation, Supervision, Writing - original draft, Writing - review \& editing. {\bf Hernán F. Astudillo:} Supervision, Writing - review  \& editing.

%==============================================================
\section{Declaration of competing interest}
The authors declare that they have no known competing financial interests or personal relationships that could have appeared to influence the work reported in this paper.

%==============================================================
\section{Acknowledgements}

This work was funded by the Fondo Nacional de Desarrollo Científico y Tecnológico (FONDECYT, CHILE) under grant FONDECYT Postdoctoral 3240226 (I.O.)

% =========================
% BIBLIOGRAPHY FOR MAIN TEXT
% =========================
\bibliographystyle{plainnat}% unsrt
\bibliography{references}

\clearpage
\appendix

% =========================
% RESET COUNTERS FOR SUPPLEMENT
% =========================
\renewcommand{\thesection}{\arabic{section}}

\setcounter{figure}{0}
\setcounter{table}{0}
\setcounter{equation}{0}

\renewcommand{\thefigure}{S\arabic{figure}}
\renewcommand{\thetable}{S\arabic{table}}
\renewcommand{\theequation}{S\arabic{equation}}

\begin{center}
    {\LARGE Supplementary Material}
\end{center}

\vspace{1em}
% =========================
% START A NEW LOCAL TOC FOR SUPPLEMENT
% =========================
\startcontents[supp]

\section*{Contents}
\printcontents[supp]{}{1}{} % Indice (Contents)
\vspace{2em}     % Espacio extra antes de empezar

%===========================================================
\newpage
\section{ODD Documentation of the Model}

The model description follows the ODD (Overview, Design concepts, Details) protocol for describing individual- and agent-based models \cite{grimm_2006}, as updated by Grimm et al. \cite{grimm_2020}.

%========== Section 1 ==========
\subsection{Purpose and Patterns}

The purpose of this model is to investigate how the composition of a parliament affects legislative efficiency. In particular, it examines how the proportion of independent legislators elected by sortition influences the approval rate of laws, average social welfare, and the overall efficiency of parliament. This model is both exploratory and explanatory: it seeks to identify behavioral thresholds that emerge as the number of independent legislators varies and to characterize the interaction between individual and collective interests.

The outcomes of interest are (1) the percentage of proposals passed, (2) the average social benefit of passed laws, and (3) overall efficiency, defined as the product of these two metrics. These patterns are derived from multiple legislatures and are used to determine whether adding independent legislators improves or worsens legislative efficiency, as well as to identify critical values for the proportion of independent legislators at which trends in law approval and welfare change.

%========== Section 2 ==========
\subsection{Entities, State Variables and Scales}

The model’s entities are agents representing legislators in a parliament. One fraction of them is affiliated with political parties and were elected by vote, and the other fraction of legislators is selected by sortition. The agents elected by sortition are called Independent legislators ($Ind$), since they are not subject to party discipline and act autonomously in evaluating legislative proposals. The agents elected by vote are affiliated with the majority party ($P_M$) or the minority party ($P_m$). Although parties are formed from legislators elected by vote, they are identified as collective entities because they have their own state variables (center and radius) and voting rules that determine the behavior of their affiliated legislators. The environment consists of a normalised Cartesian plane \cite{pluchino_2011a} between $[-1,1]$, which represents the interests that guide their behavior. The horizontal axis represents personal gain, and the vertical axis represents social gain.

The following table summarizes all the state variables of entities.

\begin{center}
\begin{tabular}{|l|p{10cm}|}
\hline
\textbf{Entity} & \textbf{State variables (brief description)} \\
\hline
Legislator ($i$) & \begin{itemize}[nosep, leftmargin=*, before=\vspace{-0.5\baselineskip}]
                        \item \textit{Position} $(x_i, y_i)$: an integer number between $[-1,1]$ that represents coordinates in the Cartesian plane; 
                        \item \textit{Affiliation} $\in\{\mathit{Ind},\mathit{P_M},\mathit{P_m}\}$; 
                        \item \textit{Personal acceptance window}: a rectangle containing all points at or above their position in the plane defined as $\{(x,y)\mid x\ge x_i\wedge y\ge y_i\}$.
                    \end{itemize} \\
\hline
Party ($k$) & \begin{itemize}[nosep, leftmargin=*, before=\vspace{-0.5\baselineskip}]
                \item \textit{Party centre} $(x_k,y_k)$: an integer number drawn uniformly between $[-1+r_k,1-r_k]$that represents coordinates in the Cartesian plane;
                \item \textit{tolerance radius} $r_k$: a real number between $[0,1]$ that defines dispersion of members and the Party acceptance window;
                \item \textit{Party acceptance window}: a rectangle containing all points at or above their position in the plane defined as $\{(x,y)\mid x\ge x_k\wedge y\ge y_k\}$.
             \end{itemize} \\
\hline
Global &  \begin{itemize}[nosep, leftmargin=*, before=\vspace{-0.5\baselineskip}]
            \item $N$: total number of legislators;
            \item $p$: proportion of legislators affiliated with the majority party;
            \item $N_{\text{ind}}$: number of independent legislators;
            \item $N_a$: number of acts per legislature;
            \item $L$: number of legislatures.
          \end{itemize} \\
\hline
\end{tabular}
\end{center}

The model represents space as a continuous plane (Cartesian plane) that captures their ideological position and determines their voting behavior. Agents remain within this domain after initial placement. So, the idea of space as a means of interaction or as a location for agents within one-, two-, or three-dimensional cells is not included.

Time scale progresses in discrete legislative acts. Each legislature consists of $N_a$ acts; after each act, relevant variables are updated. After $L$ legislatures, aggregate statistics are computed. Continuous time is not modelled.

%========== Section 3 ==========
\subsection{Process Overview and Scheduling}

The model covers the entire legislative term, with a fixed number of legislative acts, and the simulation begins after the initial conditions are established. Next, the agents' actions take place at each time step or legislative act until the end of the simulation, following this order: 1) Choose a legislator $i$ uniformly at random. 2) The selected legislator submits a proposal. 3) The proponent votes in favor of their proposal. 4) Each independent legislator votes in favor if and only if the proposal falls within their personal acceptance window. 5) Each legislator affiliated with a party evaluates whether the proponent belongs to their party. If both are from the same party, they vote in favor of the law. Otherwise, the voter uses the party's acceptance window. 6) Once all legislators have voted, the percentage of accepted proposals, the average social welfare for accepted proposals, and finally the overall efficiency are calculated.
%------------------------------
\begin{figure}[tb]
    \centering
    \includegraphics[width=0.9\textwidth]{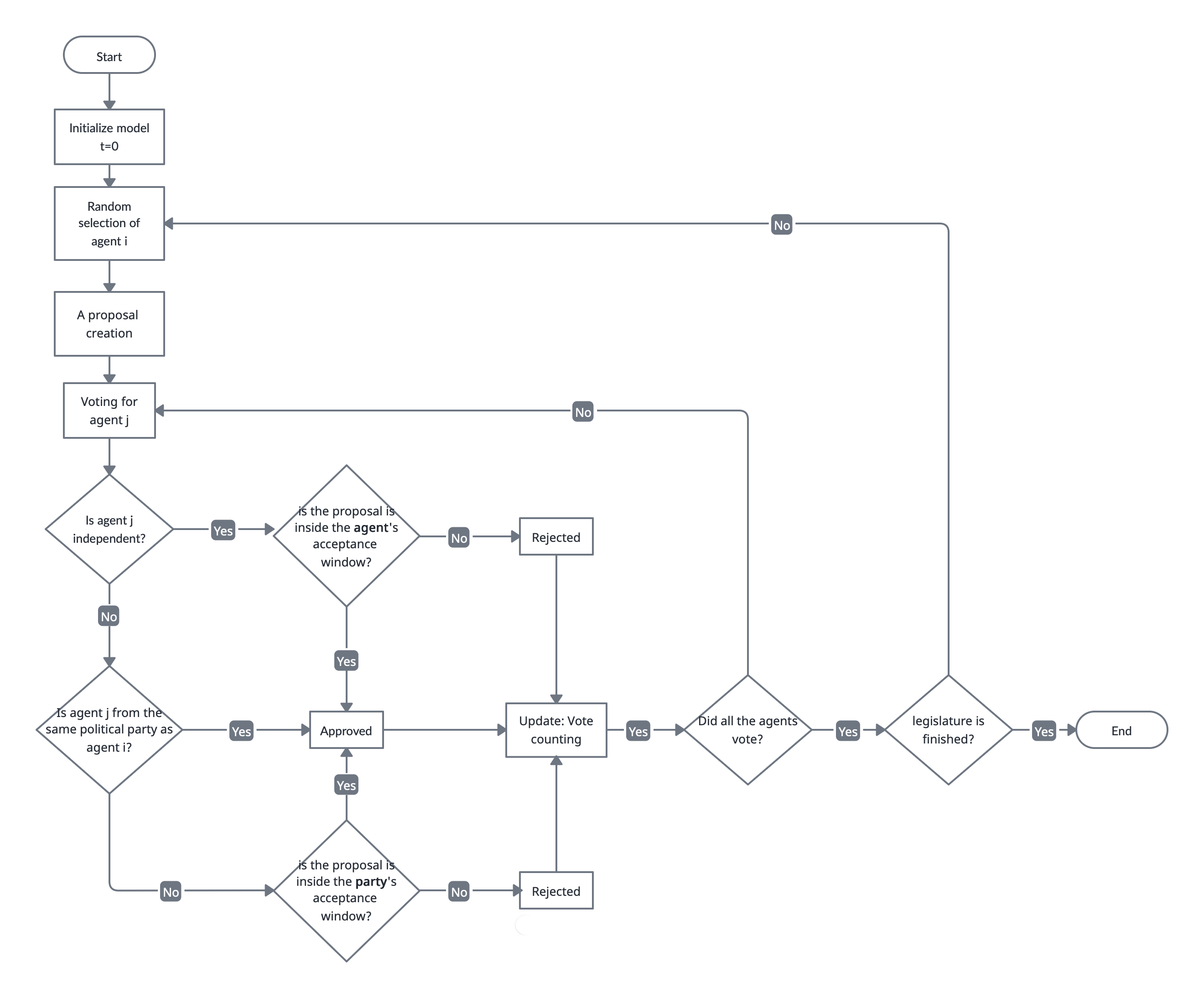}
    \caption{Flowchart of the initialization process for the parliamentary dynamics model.}
    \label{fig:initflow}
\end{figure}
%------------------------------

\noindent The flowchart can be seen in Figure~\ref{fig:initflow}, and the pseudocode for the process is as follows:

\begin{verbatim}
# INITIALISATIONS

# GLOBAL PARAMETERS
Define the global parameters:
    N       : total number of legislators
    p       : proportion of legislators affiliated with the majority party
    r       : tolerance radius for party dispersion
    q       : the quorum required to the majority rule
    N_a     : number of legislative acts per legislature
    L       : number of legislatures to simulate
    N_ind   : number of independent legislators

# PARTIES
For each party k in {PM, Pm}:
    Set the party tolerance radius r_k = r
    Draw a party centre C_k = (x_k, y_k) uniformly in [-1 + r, 1 - r]
    

# ASSIGN PARTY LEGISLATORS
    N_M = (N - N_ind) * p         # majority party members
    N_m = (N - N_ind) * (1 - p)   # minority party members
    
Create N_M legislators for the majority party and N_m for the minority party
For each party legislator i:
    Sample (x_i, y_i) from a normal distribution with mean C_k and standard deviation r/3
    Assign affiliation_i = k

# ASSIGN INDEPENDENT LEGISLATORS
Create N_ind independent legislators
For each independent legislator i:
    Sample (x_i, y_i) from independent uniform distributions with mean 0 and standard deviation 1
    Assign affiliation_i = independent

# INITIALISE COUNTERS
Set N_acc = 0            # count of accepted proposals
Set WelfareSum = 0        # cumulative social welfare

# EXECUTION OF LEGISLATURES
For L in 1 .. L:
    # EXECUTION OF ACTS
    For act in 1 .. N_a:
        # Select a proponent uniformly at random from all legislators
        Choose an index i uniformly from {1, ..., N}
        proposer = legislator_i

        # Proposal creation
        y_n = y_i                      # social benefit equals proposer’s y
        x_n = x_i                      # personal benefit for proposer

        # Perception of the proposal
        For each legislator j in 1 .. N:
            Draw x_nj uniformly from [-1, 1]    # personal benefit perceived by j
            Set y_nj = y_n                      # social benefit perceived accurately

            # Voting rules
            if j == i:
                vote_j = 1                      # proposer always votes in favour
            else if affiliation_j == independent:
                if x_nj >= x_j and y_nj >= y_j:
                    vote_j = 1
                else:
                    vote_j = 0
            else if affiliation_j == affiliation_i:
                vote_j = 1                      # automatic approval within party
            else:
                # Compare against the acceptance window of party k
                let k = affiliation_j
                if x_nj >= x_k and y_nj >= y_k:
                    vote_j = 1
                else:
                    vote_j = 0

        # Vote counting and outcome
        total_votes = sum of vote_j over all legislators
        if total_votes >= floor(qN) + 1:
            N_acc += 1
            WelfareSum += y_n

    # END OF LEGISLATURE
    Compute acceptance_rate = N_acc / N_a
    if N_acc > 0:
        Compute mean_welfare = WelfareSum / N_acc
    else:
        mean_welfare = 0
    Compute efficiency = acceptance_rate * mean_welfare
    Record acceptance_rate, mean_welfare and efficiency for legislature L

    # Reinitialise counters for next legislature
    N_acc = 0
    WelfareSum = 0

# END PROGRAM

\end{verbatim}

%========== Section 4 ==========
\subsection{Design Concepts}

Below we summarise how each concept applies to this model.

\begin{enumerate}
  \item \textit{Basic principles.} The model is grounded in the Cartesian plane, which classifies actions by their personal and social benefits. It assumes legislators seek to maximise both personal and collective gains and that party discipline influences voting. 
 \item \textit{Emergence.} Legislative efficiency and acceptance rates emerge from individual voting decisions. The nonlinear relationship between the number of independents and global efficiency is considered an emergent outcome.
 \item \textit{Adaptation.} Agents do not adapt or change strategy. Legislators maintain fixed positions in Cartesian plane throughout a legislature and do not learn; variability stems only from random proposer selection and the random perception of the personal gain of the proposal that legislators vote on.
 \item \textit{Objectives.} Legislators implicitly aim to support proposals that offer at least as much personal and social benefit as their own situation (or that of their party). Parties seek to pass proposals aligned with their principles and enforce discipline by automatically approving proposals from their members.
 \item \textit{Learning.} The model includes no learning; voting rules remain fixed across acts and legislatures.
 \item \textit{Prediction.} Legislators do not explicitly predict future outcomes; decisions are based solely on the current proposal and acceptance criteria.
 \item \textit{Sensing.} Legislators perceive the social benefit of a proposal they must vote on exactly, and know their own coordinates or those of their party. The personal benefit that drives the proposal is randomly assigned to each voter, reflecting individual interpretations of personal gain. 
 \item \textit{Interaction.} Interaction occurs through voting and party discipline. Decisions are combined through summation of votes; there is no negotiation or communication during a legislative act. Party affiliation creates an indirect interaction by delegating some decision‑making to the party acceptance window.
 \item \textit{Stochasticity.} The model contains several sources of randomness: the initial positions of legislators and party centres; the selection of proposers; and the random draw of each voter's personal benefit from the proposal. This stochasticity introduces variability and allows the exploration of distributions of outcomes.
 \item \textit{Collectives.} Parties are explicit collectives with their own state variables (centre and radius). Members automatically approve proposals from their own party and use a collective acceptance window for external proposals. Independents do not form additional collectives.
 \item \textit{Observation.} Key outputs recorded at the end of each legislature are the percentage of proposals accepted, the mean social welfare of accepted proposals, the global efficiency and the average number of favourable votes.
\end{enumerate}

%========== Section 5 ==========
\subsection{Initialization}

Initialization creates a new legislation, establishes global parameters, defines all entities, and finally defines tracking variables. The following explicitly describes all the steps necessary to configure the model.

\begin{enumerate}
 \item \textbf{Global parameters.} First, set the parliament size $N$, the percentage of legislators of the majority party $p$, the number of independent legislators $N_{ind}$, the number of legislative acts $N_a$, the legislatures $L$, the tolerance radius $r$, and the majority rules quorum $q$. All the simulations used in this work are: $N = 1000$, $p \in \{0.51, 0.6, 0.8\}$, $N_{ind} \in [0, 1000]$ , $N_a = 1000$, $L = 1000$, $r_k = r = 0.2$ and $q \in \{1/2, 4/7, 3/5, 2/3\}$.

 \item \textbf{Party creation.} For each party $k\in\{\mathit{P_M},\mathit{P_m}\}$, select a centre $C_k=(x_k,y_k)$, and their location is generated using a uniform distribution on both axes in the interval $[-1+r_k,1-r_k]$, following the method presented by Caserta et al. \cite{caserta_2021}. These coordinates remain fixed during a legislature.
  
 \item \textbf{Assign party legislators.} First, compute the number of legislators affiliated with the majority party $N_M$ and the number of legislators affiliated with the minority party $N_m$ using the following formulas: $N_M = (N - N_{ind})p$ and $N_m = (N - N_{ind}) (1 - p)$. Then, the placement of legislators belonging to each party is performed using a normal distribution on the $x$ and $y$ axes, with means $x_k$ and $y_k$, respectively, and standard deviations $r_k/3$. These set values give a probability of $99.73\%$ that the random value falls within the circle of radius $r$ \cite{pukelsheim_1994}.
 
 \item \textbf{Assign independent legislators.} Independent legislators $N_{\text{ind}}$ are randomly placed anywhere on the diagram using a normal distribution on both axes and centred at zero for each coordinate. Independents thus spread across the Cipolla plane. 

 \item \textbf{Initialise tracking variables.} Set counters for accepted laws $N_{\text{acc}}$ and the accumulated social welfare to zero. Votes do not carry over between acts.
\end{enumerate}

%========== Section 6 ==========
\subsection{Input Data}

The model uses no external time series or exogenous data. All inputs are fixed parameters chosen before running simulations; thus no additional input data are needed.

%========== Section 7 ==========
\subsection{Submodels}

The main submodels are:

\begin{enumerate}
\item \textbf{Proposal generation}
    \begin{enumerate}
        \item[] \textit{Input:} list of legislators with positions and affiliations.
        \item[] \textit{Process:} select a legislator $i$ at random and set $y_n := y_i$. Set $x_n := x_i$ for the proposer; this coordinate records the proposer’s personal benefit.
        \item[] \textit{Output:} the proposer's party affiliation and the proposal $a_n=(x_n,y_n)$.
    \end{enumerate}

\item \textbf{Assignment of perceptions}
    \begin{enumerate}
        \item[] \textit{Input:} proposal $a_n=(x_n,y_n)$ and the set of legislators.
        \item[] \textit{Process:} for each legislator $j$, draw $x_{nj}\sim U(-1,1)$, an independent uniform random number in $[-1,1]$. Set $y_{nj}=y_n$ (the social benefit is perceived accurately).
        \item[] \textit{Output:} perceived pairs $(x_{nj},y_n)$ used to evaluate the proposal.
    \end{enumerate}

\item \textbf{Acceptance windows and voting}
    \begin{enumerate}
        \item[] \textit{Input:} perceived values $(x_{nj},y_n)$ and state variables of legislators and parties.
        \item[] \textit{Process:}
            \begin{itemize}
                \item For each independent legislator $j$, vote yes if $x_{nj}\ge x_j$ and $y_n\ge y_j$, otherwise vote no.
                \item For each party legislator $j$:
                    \begin{itemize}
                        \item If $j$ and the proposer belong to the same party, approve automatically.
                        \item Otherwise, use the party acceptance window: vote yes if $x_{nj}\ge x_k$ and $y_n\ge y_k$, and no otherwise.
                    \end{itemize}
            \end{itemize}
        \item[] \textit{Output:} individual votes $v_j\in\{0,1\}$.
    \end{enumerate}

\item \textbf{Vote counting and metric updates}
    \begin{enumerate}
        \item[] \textit{Input:} list of votes $v_j$, value $y_n$, and global counters.
        \item[] \textit{Process:} compute $V = \sum_j v_j$. If $V\ge qN + 1$, increment $N_{\text{acc}}$ and add $y_n$ to the accumulated social welfare; otherwise discard the proposal.
        \item[] \textit{Output:} updated values of $N_{\text{acc}}$ and total welfare.
    \end{enumerate}

\item \textbf{Indicator calculation}
    \begin{enumerate}
        \item[] \textit{Input:} at the end of each legislature: $N_{\text{acc}}$, accumulated welfare and number of acts $N_a$.
        \item[] \textit{Process:} compute $N\%_{\text{acc}}(L) = N_{\text{acc}}/N_a$; compute $Y(L)$ as the average of $y_n$ for accepted proposals; compute the global efficiency $\mathrm{E\!f\!f}(L) = N\%_{\text{acc}}(L) \times Y(L)$.
        \item[] \textit{Output:} indicators $N_{\%\text{acc}}$, $Y(L)$ and $\mathrm{E\!f\!f}(L)$; mean number of favourable votes.
    \end{enumerate}
\end{enumerate}

%===========================================================
\newpage
\section{Details of the model and Additional results}

\subsection{The ideological Cartesian plane}

The model presented by Pluchino et al. \cite{pluchino_2011a} investigates how the composition of a parliament affects legislative efficiency. To do so, it introduces agents who vote on proposals based on their ideological positions and personal interests. To formally represent the principles and values that guide each agent's behavior, the model places them on a two-dimensional Cartesian plane, normalized to the interval $[-1, 1]$ on both axes.

On this plane, the horizontal axis ($p$) represents the personal benefit associated with a legislative proposal, while the vertical axis ($q$) represents the social benefit derived from that proposal. In this way, each point $(p, q)$ simultaneously expresses the individual and collective evaluations of a legislative action, allowing the agent's ideological orientation to be characterized synthetically.

\begin{figure}
    \centering
    \includegraphics[width=0.50\linewidth]{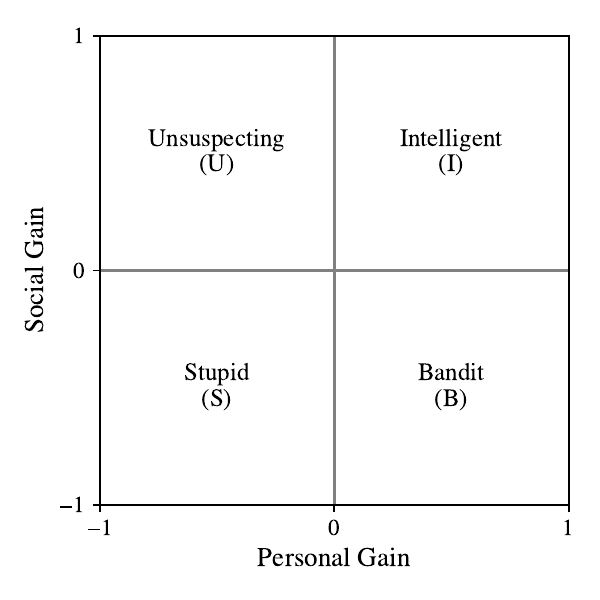}
    \caption[Cipolla diagram]{\label{fig_cipolla_diagram} Ideological Cartesian plane based on the diagram presented by Carlo M. Cipolla in his essay on the fundamental laws of human stupidity \cite{cipolla_2021}.}
\end{figure}

This scheme is inspired by the diagram proposed by Carlo M. Cipolla in his essay on the fundamental laws of human stupidity \cite{cipolla_2021}. In this diagram (\hyperref[fig_cipolla_diagram]{Fig.~\ref{fig_cipolla_diagram}}), individual actions are classified into four categories according to the sign of their personal and social effects: intelligent ($I$), whose actions generate both individual and collective benefits; unsuspecting ($U$), who produce social benefits even when they involve a personal cost; bad ($B$), who obtain personal gains at the expense of social harm; and stupid ($S$), whose actions generate losses for both themselves and society.

Mathematically, if $p$ denotes personal gain and $q$ denotes social gain, the conditions that delimit each region of the plane are:
\begin{align*}
    S &: p \leq 0 \text{ and } q < 0 \\
    U &: p \leq 0 \text{ and } q \geq 0 \\
    I &: p > 0 \text{ and } q \geq 0 \\
    B &: p > 0 \text{ and } q < 0.
\end{align*}

In the context of the parliamentary model, this plane does not seek to classify legislators in an essentialist manner, but rather to represent, operationally, the average orientation of their decisions. Given that individuals may act inconsistently over time, each agent's position on the plane should be interpreted as a weighted average of their legislative actions, capturing the dominant trend of their behavior rather than each decision. Thus, the ideological Cartesian plane functions as a formal tool that translates personal motivations and social impacts into a simple geometric representation, facilitating the aggregate analysis of parliamentary dynamics within the model.

%------------------------------
%------------------------------

%===========================================================
\subsection{Additional results and robustness analysis}

To systematically characterize the model's behavior and evaluate the robustness of the main results, we conducted a controlled parametric exploration of the system. This exploration allows us to identify the sensitivity of macroscopic metrics to variations in system size, required quorum, and party composition.

Unless otherwise indicated, we kept the following parameters constant: the tolerance radius of the parties, set at $r_k = r = 0.2$; the number of legislative acts per legislature, $N_a = 1000$; and the number of legislatures used to calculate averages, $N_L = 1000$. These values ensure adequate statistical convergence of the macroscopic metrics and reduce the noise associated with finite fluctuations. The number of independent legislators, $N_{ind}$, was systematically varied in steps of $2$, generating configurations with $N_{ind} = 1, 3, 5, \dots, 1000$. This procedure allows us to exhaustively cover the entire range of possible parliamentary compositions and analyze the transition between different collective regimes.

\begin{figure}
    \centering
    \includegraphics[width=0.65\linewidth]{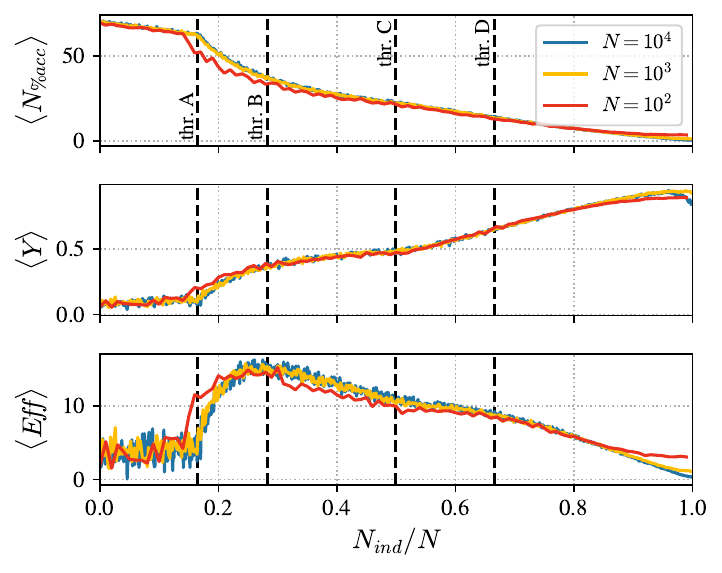}
    \caption[fig s4]{\label{fig:size_variation} Finite-size scalability. Average behavior of macroscopic metrics for different system sizes. The curves preserve their qualitative structure as $N$ increases. Parameters used: $r = 0.2$, $N_L = 1000$, $N_a = 1000$, $p = 0.6$, and $q = 0.5$.}
    \
\end{figure}

The first parameter that we explored was the total size of the parliament, considering $N = [10^2, 10^3, 10^4]$. Fig.~\ref{fig:size_variation} shows that the macroscopic metrics qualitatively retain their structure as the system size increases, while the statistical noise decreases progressively. In particular, the transitions between regimes and the relative positions of the thresholds remain invariant, indicating that the observed behavior is not a finite-size artifact. This result demonstrates the model's scalability and enables us to select an intermediate size that balances resolution and computational cost. Consequently, we use $N = 1000$ for the main results of the study.

%------------------------------
%------------------------------

\begin{figure}
    \centering
    \includegraphics[width=0.65\linewidth]{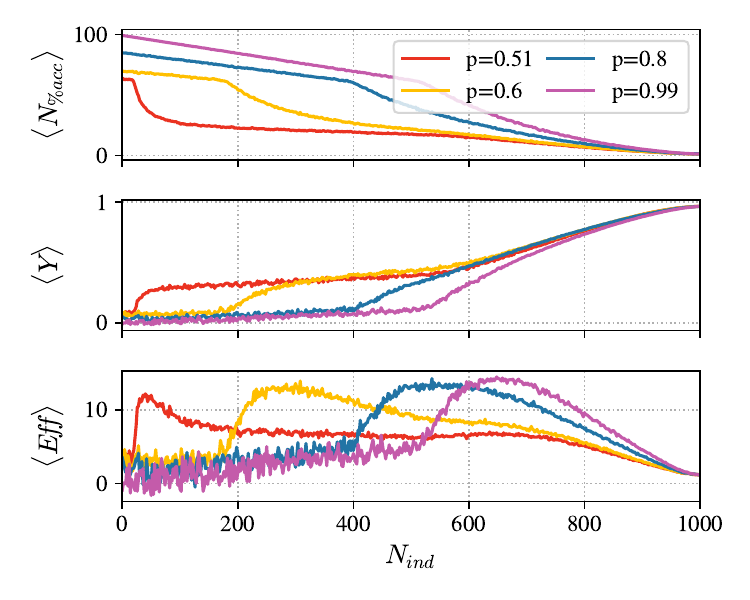}
    \caption[fig s4]{\label{fig:variacion_p} Variation of macroscopic metrics according to the proportion of legislators from the majority party. Parameters used: $r = 0.2$, $N_L = 1000$, $N_a = 1000$, $q = 0.5$ and $N = 1000$.}
\end{figure}

Then, we explored the effect of the proportion of legislators belonging to the majority party, considering $p = [0.51, 0.6, 0.8, 0.99]$, where $p$ denotes the fraction of total party legislators belonging to the majority party. The results are shown in Fig.~\ref{fig:variacion_p}.

Figure~\ref{fig:variacion_p} shows that, with a fixed quorum $q=1/2$, the party proportion $p$ systematically reconfigures the balance between approval and social welfare as $N_{ind}$ increases. In the upper panel, larger majorities ($p=0.8$ and $p=0.99$) maintain high approval rates across a wider range of $N_{ind}$, whereas for $p=0.51$, approval falls sharply even with few independents, indicating greater coalition fragility. In the middle panel, $\langle Y \rangle$ increases with $N_{ind}$ in all cases. However, for high $p$, this increase is delayed: the system approves "typical" proposals with low social selectivity for longer, and only when independents reach a critical mass does the set of viable proposals change, and $\langle Y \rangle$ rises sharply. As a result, efficiency $\langle E\!f\!f \rangle$ exhibits a non-monotonic maximum whose optimum point shifts toward increasing values of $N_{ind}$ as $p$ increases, reflecting the trade-off between the number of laws passed and their average social quality. Taken together, these patterns are consistent with the microscopic interpretation discussed above: increasing $p$ expands the majority party's ability to pass laws, but also delays the "corrective" effect of independents on social welfare, shifting the maximum efficiency toward more heterogeneous parliaments.

%------------------------------
%------------------------------

%===========================================================
\subsection{Analytical determination of thresholds}

In this section, we describe in detail the analytical approximations used to obtain the thresholds that delimit the different regions of the phase diagram. The analysis is based on a mean-field approximation, which aims to derive the macroscopic behavior of the system from expected values, while ignoring finite fluctuations and microscopic correlations between agents.

The mean-field approximation replaces the specific configuration of individual positions with their average behavior in the limit of a large number of legislators. In particular, given the design of the model, both the positions of independent legislators and the ideological centers of the parties are uniformly distributed in the normalized Cartesian plane in the interval $[-1,1]$. On the other hand, the individual positions of independents are statistically independent, implying no spatial correlations or strategic coordination among them. So that their aggregate behavior can be treated as a statistical superposition, and the expected number of favorable votes is the sum of their individual expectations. Finally, collective behavior can be approximated by the expected number of favorable votes, disregarding fluctuations of the order $1/\sqrt{N}$. By disregarding fluctuations, it is assumed that in sufficiently large parliaments the effective number of votes is concentrated around its expected value (law of large numbers). Consequently, the approval of proposals can be analyzed deterministically in terms of average values, which allows the thresholds that delimit the different regions of the phase diagram to be derived analytically.\newline

For a legislative proposal, the horizontal coordinate (personal benefit) is uniformly distributed in the interval $[-1,1]$, while the vertical coordinate (social benefit) coincides with the vertical position of the proposing legislator. Since both independents and party members are uniformly distributed on the vertical axis, the expected value of the social benefit coordinate of the proposals is zero. Likewise, uniformity on the horizontal axis implies that the expected value of personal benefit is also zero. Therefore, under the mean-field approximation, the average legislative proposal lies at the center of the ideological plane, i.e., at $(0,0)$. This simplification allows us to analyze the expected voting structure without considering the complete distribution of proposals.

\begin{figure}
    \centering
    \includegraphics[width=0.50\linewidth]{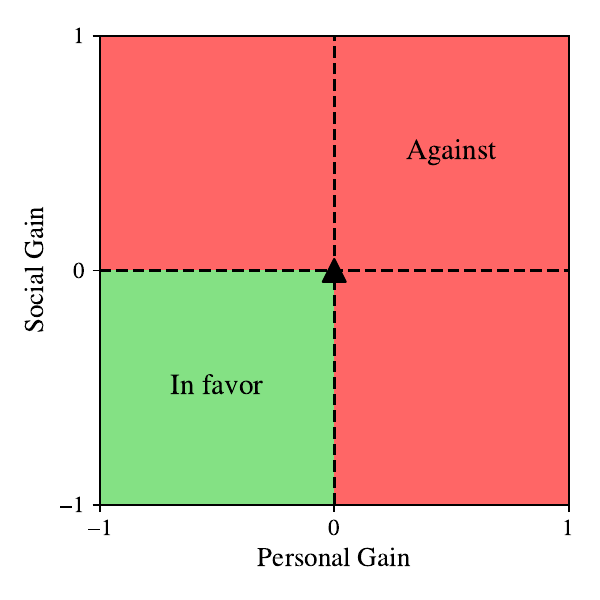}
    \caption[fig s5]{\label{fig_s2} Diagram showing the average position of the proposed legislation ($\blacktriangle$). Independent legislators in the lower-left quadrant tend to vote in favor, while those in the other quadrants tend to vote against. Elaborated based on the presentation by Pluchino et al. \cite{pluchino_2011a}.}
\end{figure}

Given that independent legislators are evenly distributed across the plane, the probability of an independent being located in any of the four quadrants is $1/4$. Under the model's decision rule, an independent will tend to vote in favor of the average proposal if their position is in the lower-left quadrant, that is, when both their expected personal and social benefits with respect to the average proposal are negative. Consequently, the expected number of independents voting in favor of an average proposal is equal to one-quarter of the total number of independents. Figure \ref{fig_s2}  shows a schematic representation of this estimate. On average, then, the independent support bloc can be approximated by the fraction $N_{ind}/4$, where $N_{ind}$ is the total number of independent legislators.

%------------------------------
%------------------------------

Let $N$ be the total number of legislators, composed of $N_M$ members of the majority party, $N_m$ of the minority party, and $N_{ind}$ independents, such that $N = N_M + N_m + N_{ind}$ and $N_M = (N - N_{ind})p$, with $p$ the proportion that ensures the majority. Let $q$ be the quorum under the majority rule considered. Under the mean field approximation, the expected number of votes in favor of an average proposal depends on the parliamentary configuration. Then, the thresholds are obtained as follows:

\begin{enumerate}
    \item Suppose $N_M \geq q N+1$; then the majority party can pass proposals without external support. Later the first threshold is,
        \begin{align}
            N_M &= q N+1 \nonumber \\
            (N - N_{ind}^A)p &= q N+1 \nonumber\\
            \text{thr. A:} \quad N_{ind}^A &= \frac{N(p-q) - 1}{p}.
        \end{align}
    
    \item When $N_M < q N+1$ but $N_M + N_{ind}/4 \geq q N+1$, the majority party requires the expected support of a quarter of the independents to reach a quorum. Then, the second threshold is,
        \begin{align}
        N_M + \frac{N_{ind}^B}{4} &= qN + 1 \nonumber\\
        (N - N_{ind}^B)p + \frac{N_{ind}^B}{4}&=qN + 1 \nonumber\\
        N_{ind}^B(p-0.25) &=Np - qN - 1 \nonumber\\
        \text{thr. B:} \quad N_{ind}^B &= \frac{N(p-q)-1}{p-0.25}.
    \end{align}
    
    \item If the above condition is not met, but $N_M + N_m \geq q N+1$, i.e., a coalition between both parties can approve the proposal. The third threshold is, 
        \begin{align}
        N_M + N_m &= qN + 1 \nonumber\\
        N - N_{ind}^C &= qN + 1 \nonumber\\
        \text{thr. C:} \quad N_{ind}^C &= N(1-q) - 1 .
    \end{align}
    
    \item Finally, when none of the above configurations satisfies the quorum, the joint support of the majority party, the minority party, and the favorable fraction of independents is necessary, that is, $N_M + N_m + N_{ind}/4 \geq q N+1$. So, the fourth threshold is, 
        \begin{align}
        N_M + N_m + \frac{N_{ind}^D}{4} &= qN + 1 \nonumber\\
        N - N_{ind}^D + \frac{N_{ind}^D}{4} &= qN + 1 \nonumber\\
        \frac{3}{4}N_{ind}^D &= N (1-q) - 1 \nonumber\\
        \text{thr. D:} \quad N_{ind}^D &= \frac{4}{3}N(1-q) - \frac{4}{3} .
    \end{align}
\end{enumerate}

\begin{figure}
    \centering
    \includegraphics[width=0.60\linewidth]{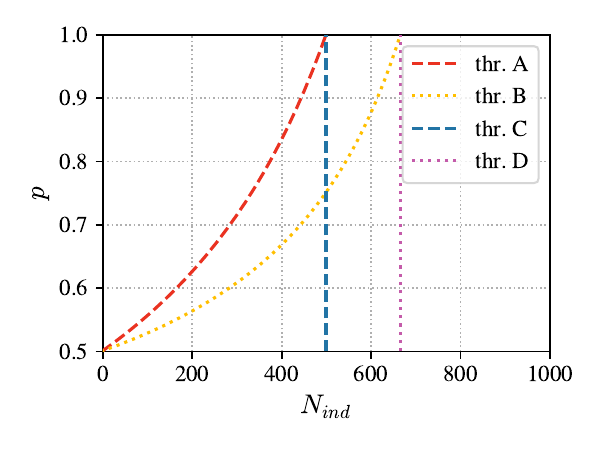}
    \caption[fig s6]{\label{fig_theo_phase_dia}Theoretical Phase Diagram. Thresholds calculated analytically as a function of $N_{ind}$ and $p$ for $q=0.5$.}
\end{figure}

Based on these thresholds, the phase diagrams allow us to identify the microscopic dynamics underlying each collective regime. For example, Figure \ref{fig_theo_phase_dia} shows a phase diagram for the quorum $q=1/2=0.5$, used in the simple majority rule.

%------------------------------
%------------------------------

%===========================================================

\subsection{Numerical determination of thresholds}

Below, we describe the numerical procedure used to estimate the analytically predicted thresholds from simulation results. In all cases, we determined thresholds as values of $N_{ind}$ associated with structural features of macroscopic curves (abrupt changes in slope, global maxima, or changes in concavity), and we estimated uncertainty using resampling schemes when appropriate.

\subsubsection*{Threshold A}

To numerically determine threshold A, we use the curve of approved proposals, $\langle N_{\%acc} \rangle$, which shows a sudden change in trend compatible with an effective discontinuity in the local slope. To locate this point, we estimate local slopes by linear regression in windows to the left and right of each point. In particular, for each index $i$ we consider the intervals $N_{ind}[i-w,i]$ and $N_{ind}[i,i+w]$, where $w$ is the window size, and we calculate the slopes $D_{\text{atr.}}$ and $D_{\text{ade.}}$ using a linear fit (NumPy \texttt{polyfit}), along with their standard errors $\sigma_{\text{backw.}}$ and $\sigma_{\text{forw.}}$.

We define the break magnitude as

$$
\Delta D = |D_{\text{backw.}} - D_{\text{forw.}}|,
$$

and its associated uncertainty

$$
\sigma_{\Delta} = \sqrt{\sigma_{\text{backw.}}^2 + \sigma_{\text{forw.}}^2}.
$$

Then, we identified the threshold A with the value of $N_{ind}$ where $\Delta D$ reaches its global maximum. To estimate the associated uncertainty, we consider the set of points that satisfy,

$$
\Delta D \ge \Delta D_{\max} - \sigma_{\Delta}.
$$

\begin{figure}
    \centering
    \includegraphics[width=0.65\linewidth]{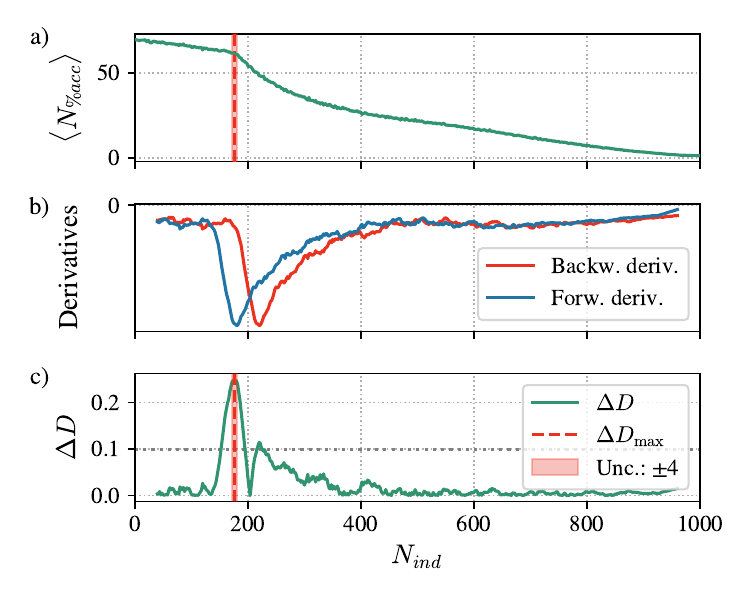}
    \caption[fig s7]{Example of the numerical procedure used to identify the Threshold A. (a) Average percentage of accepted proposals, $\langle N_{\%acc} \rangle$, as a function of $N_{ind}$, calculated over $N_L = 1000$ legislatures with fixed parameters ($r = 0.2$, $N_a = 1000$, $p = 0.6$, $q = 0.5$). (b) Backward and forward derivatives estimated using linear adjustments in windows of size $w=20$ points backward and forward, respectively. (c) Absolute difference $\Delta D = |D_{\text{backw.}} - D_{\text{forw.}}|$, indicating the global maximum $\Delta D_{\text{max}}$, whose position defines the breakpoint ($N_{ind}=176$ in this case), and the interval used to estimate its uncertainty ($\pm 4$). The dotted line at $\Delta D=0.1$ represents the minimum threshold required to consider the break statistically significant.}
    \label{fig:umbral_a}
\end{figure}

Furthermore, we define the error as the half-width of the interval of $N_{ind}$ covered by that set. Fig.~\ref{fig:umbral_a} shows an illustration of the procedure.

%------------------------------
%------------------------------

\subsubsection*{Threshold B}

To estimate threshold B, we identify the point of maximum efficiency $\langle E\!f\!f \rangle$ as a function of $N_{ind}$ \cite{pluchino_2011a}. Since the curve may contain statistical noise, we first smooth it using a Savitzky–Golay filter \cite{savitzky_golay_1964}, implemented in SciPy, using an odd window of 71 points and a third-order polynomial.

\begin{figure}
    \centering
    \includegraphics[width=0.65\linewidth]{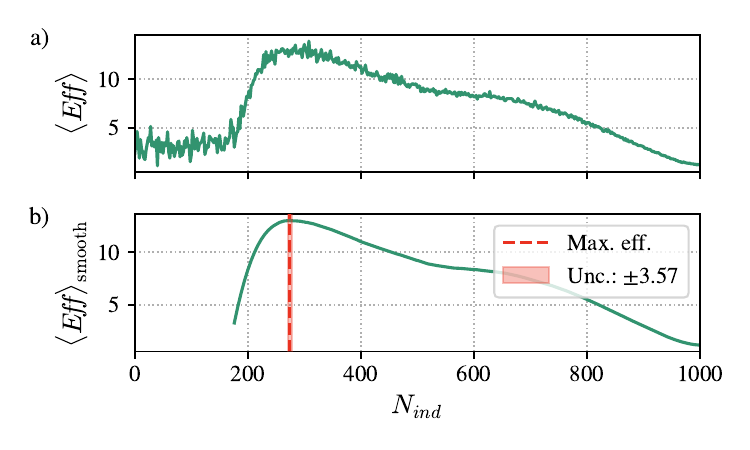}
    \caption[fig s8]{Example of the numerical procedure used to determine threshold B. (a) Average efficiency of parliament, $\langle E\!f\!f \rangle$, as a function of $N_{ind}$, calculated over $N_L = 1000$ legislatures with fixed parameters ($N_a = 1000$, $r = 0.2$, $p = 0.6$, $q = 0.5$). (b) Smoothed curve $\langle E\!f\!f \rangle_{\text{smooth}}$ obtained using the Savitzky–Golay filter (window of $71$ points and polynomial of order $3$), indicating the global maximum that defines point B. The associated uncertainty was estimated using parametric bootstrapping with $1000$ resamples generated from the residual with respect to the smoothed curve; the standard deviation of the maximum positions defines the reported error.}
    \label{fig:umbral_b}
\end{figure}

We define threshold B as the global maximum of the smoothed curve, $\langle E\!f\!f \rangle_{\text{smooth}}$. When threshold A is identifiable, we restrict the search to the domain $N_{ind} \ge N_A$ in order to avoid irregularities associated with the breakpoint. Figure .~\ref{fig:umbral_b} illustrates the procedure.

The uncertainty of B is estimated using a parametric bootstrap \cite{efron_tibshirani_1994}. We calculate the residual,

$$
r = \langle E\!f\!f \rangle - \langle E\!f\!f \rangle_{\text{smooth}},
$$

and its standard deviation $\sigma_r$. We then generate 1000 synthetic realizations by adding Gaussian noise $\eta \sim \mathcal{N}(0,\sigma_r)$ to the smoothed curve, and for each realization we repeat the smoothing and maximum search. The standard deviation of the positions obtained defines the uncertainty of $N_B$.

%------------------------------
%------------------------------

\subsubsection*{Thresholds C and D}

To determine thresholds C and D, we analyze concavity changes in the average social gain curve $\langle Y \rangle$, which manifest as sign changes in the second derivative with respect to $N_{ind}$. Because derivative estimation amplifies noise, we use the Savitzky–Golay filter to robustly compute derivatives. Thus, we first use a 71-point window and a polynomial of order 2 to estimate the second derivative. We then smooth with a second application of the filter using a window of 11 points and a polynomial of order 2.

When threshold A is identifiable, the analysis is performed in the domain $N_{ind} \ge N_A$ to avoid effects associated with effective non-differentiability at this threshold. In cases where A cannot be determined robustly (e.g., for high quorums), we used the entire curve.

We identified the candidates for C and D as the zero crossings of the smoothed second derivative. To select the most structurally relevant changes and avoid spurious crossings, we take as a reference the main maximum (in absolute value) of the second derivative and choose the zero crossings immediately to its left and right, thus defining points C and D.

\begin{figure}
    \centering
    \includegraphics[width=0.65\linewidth]{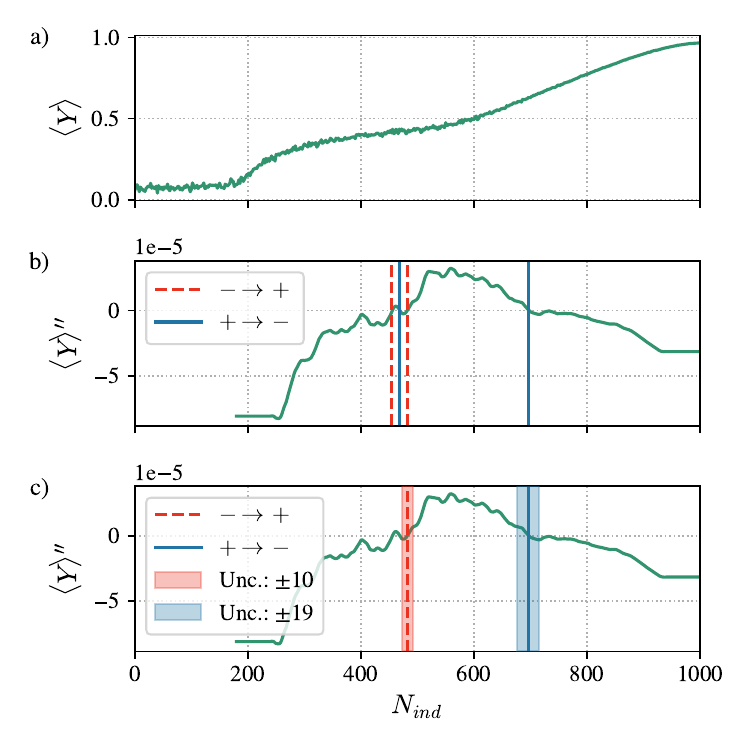}
    \caption[fig s9]{Example of the numerical procedure used to identify thresholds C and D. (a) Metric $\langle Y \rangle$ as a function of $N_{ind}$, averaged over $N_L = 1000$ legislatures with fixed parameters ($N_a = 1000$, $r = 0.2$, $p = 0.6$, $q = 0.5$). (b) Second derivative estimated after using the Savitzky–Golay filter (window of $71$ points and polynomial of order $3$), followed by additional smoothing with a window of $11$ points to reduce numerical fluctuations. The points where the second derivative changes sign are indicated. (c) Selection of the most relevant zero crossings, defined as those immediately to the left and right of the global maximum of the second derivative, which determine points C and D, together with the intervals used to estimate their uncertainties using bootstrap with $1000$ resamples.}
    \label{fig:umbral_c_d}
\end{figure}

The uncertainty of these thresholds is estimated using the same bootstrap procedure described for threshold B: the residual with respect to the smoothed curve is calculated, 1000 realizations with Gaussian noise of standard deviation equal to that of the residual are generated, and the determination of C and D is repeated in each case. The standard deviation of the positions obtained corresponds to the associated uncertainty. Fig.~\ref{fig:umbral_c_d} shows the complete procedure.

%------------------------------
%------------------------------

\end{document}